\documentclass[usenatbib]{aa}
\usepackage[varg]{txfonts}

\usepackage[T1]{fontenc}
\usepackage{ae,aecompl}

\usepackage{graphicx}
\graphicspath{{./Figures/}}
\usepackage{amsmath}    
\usepackage{amssymb}    
\usepackage{mathtools}
\usepackage{array}
\usepackage{color}

\usepackage[colorlinks=true,
linkcolor=blue,
citecolor=blue,
urlcolor=blue,
plainpages=true        
]{hyperref}     %
\usepackage{amsmath}    %
\usepackage{amssymb}    %
\usepackage{mathtools}
\usepackage{array}
\usepackage{empheq}
\usepackage{bm}%

\definecolor{darkgreen}{RGB}{0,200,0}

\renewcommand{\epsilon}{\varepsilon}                                                        

\newcommand{\ie}{\emph{i.e.}}
\newcommand{\eg}{\emph{e.g.}}

\newcommand{\Ms}{\ensuremath{\mathrm{M}_\odot}}

\renewcommand{\i}{\iota}

\newcommand{\EXP}{\mathbf{e}} %

\newcommand{\lapc}[3]{b^{(#2)}_{#1}(#3)}

\newcommand{\AMD}{\ensuremath{C}}

\newcommand{\AM}{\ensuremath{G}}
\newcommand{\CAM}{\ensuremath{\Upsilon}}
\newcommand{\scalefactor}{\ensuremath{\Gamma}}
\newcommand{\resact}{\Theta}

\newcommand{\Tsurv}{\ensuremath{T_{\mathrm{surv}}}}
\newcommand{\theres}{\ensuremath{\theta_\mathrm{res}}}
\newcommand{\hk}{h_1^{(\mathbf{k})}}

\newcommand{\smallvar}[1]{\d#1}

\newcommand{\perrat}[1]{\ensuremath{\nu_{#1}}}
\newcommand{\perratvar}[1]{\ensuremath{\d\nu_{#1}}}
\newcommand{\genperrat}{\ensuremath{\nu}}

\newcommand{\rescoefpq}{R_{pq}}

\newcommand{\plsep}{\delta}
\newcommand{\resloc}{\ensuremath{\eta}}
\newcommand{\freqpq}{\omega_{pq}}

\newcommand{\Mfac}{\ensuremath{M}}

\newcommand{\reswidth}[1]{\ensuremath{\Delta #1}}
\newcommand{\reseta}{\reswidth{\resloc_{pq}} }
\newcommand{\numfacfreq}{A}
\newcommand{\numfacres}{6.55}

\newcommand{\density}{\ensuremath{\rho}}
\newcommand{\denstot}{\ensuremath{\density_{\mathrm{tot}}}}

\newcommand{\resind}{\ensuremath{k}}

\newcommand{\plsepov}{\ensuremath{\plsep_{\mathrm{ov}}}}

\newcommand{\ovind}{\ensuremath{k_{\mathrm{ov}}}}
\newcommand{\xiov}{\ensuremath{\xi_{\mathrm{ov}}}}

\newcommand{\diffcoef}[1]{\ensuremath{D_{#1}}}
\newcommand{\diffcoefeff}{\diffcoef{\mathrm{eff}}}
\newcommand{\etadist}{\resloc_d}

\def\nnb{\nonumber\\}
\def\d{\mathrm{d}}
\def\O{\mathrm{O}}

\def\deriv#1#2{\frac{\mathrm{d} #1}{\mathrm{d} #2}}

\def\dpart#1#2{\frac{\partial #1}{\partial #2}}

\def\N{\mathbb{N}}
\def\Z{\mathbb{Z}}

\def\H{\mathcal{H}}
\def\K{\mathcal{K}}

\def\Gr{\mathcal{G}}

\usepackage[normalem]{ulem}

\graphicspath{{./Figures/}}

\begin{document}

\title{The path to instability in compact multi-planetary systems}

\subtitle{}

\author{Antoine C. Petit \inst{1,2}\and
       Gabriele Pichierri\inst{3}\and
       Melvyn B. Davies\inst{1}\and
       Anders Johansen\inst{1}
       }

\institute{Lund Observatory, Department of Astronomy and Theoretical Physics, Lund University, Box 43, 22100 Lund, Sweden\\
           \email{antoine.petit@astro.lu.se}
             \and
             IMCCE, CNRS-UMR8028, Observatoire de Paris, PSL University, Sorbonne Universit\'e, 77 Avenue Denfert-Rochereau, 75014 Paris, France
             \and
             Max-Planck-Institut f\"ur Astronomie, K\"onigstuhl 17, 69117 Heidelberg, Germany
          }

\date{}
\date{Accepted XXX. Received YYY; in original form ZZZ}

\abstract{
	The dynamical stability of tightly packed exoplanetary systems remains poorly understood.
        While a sharp stability boundary exists for a two-planet system, numerical simulations of three-planet systems and higher show that they can experience instability on timescales up to billions of years. 
	Moreover, an exponential trend between the planet orbital separation measured in units of Hill radii and the survival time has been reported. 
	While these findings have been observed in numerous numerical simulations, little is known of the actual mechanism leading to instability. 
	Contrary to a constant diffusion process, planetary systems seem to remain dynamically quiescent for most of their lifetime before a very short unstable phase. 
	In this work, we show how the slow chaotic diffusion due to the overlap of three-body resonances dominates the timescale leading to the instability for initially coplanar and circular orbits.
        While the last instability phase is related to scattering due to two-planet mean motion resonances (MMRs), for circular orbits the two-planets MMRs are too far separated to destabilise systems initially away from them.
        The studied mechanism reproduces   the qualitative behaviour found in numerical simulations very well.
        We develop an analytical model to generalise the empirical trend obtained for equal-mass and equally spaced planets to general systems on initially circular orbits.
        We obtain an analytical estimate of the survival time consistent with numerical simulations over four orders of magnitude for the planet-to-star-mass ratio $\epsilon$, and 6 to 8 orders of magnitude for the instability time.
	We also confirm that measuring the orbital spacing in terms of Hill radii is not adapted and that the right spacing unit scales as $\epsilon^{1/4}$.
	We predict that beyond a certain spacing, the three-planet resonances are not overlapped, which results in an increase of the survival time.
	We confirm these findings with the aid of numerical simulations of three-planet systems with different masses.
        We finally discuss the extension of our result to more general systems, containing more planets on initially non-circular orbits.
}


\maketitle
\section{Introduction}

One of the most astonishing results of the \emph{Kepler} mission has been the discovery of very compact super-Earth multiplanetary systems \citep{Borucki2011,Fabrycky2014}.
These systems, such as Kepler-11 \citep{Lissauer2011}, can host more than six planets with masses between that of the Earth and Neptune, and with periods of less than 100 days.
They have very low mutual inclinations and eccentricities \citep{Johansen2012,Fang2012,Xie2016} and for the majority, they are not in resonant chains \citep{Lissauer2011a,Fabrycky2014}.
Understanding the orbital properties of these so-called super-Earths or mini-Neptunes is crucial, as it seems that at least 50\% of solar-type stars host a close-in planet with a radius comprised between that of Earth and Neptune \citep{Mayor2011,Petigura2013,Fressin2013}.

Studies of the \emph{Kepler} multiplanetary systems have shown that the architecture is most likely sculpted by dynamical stability \citep{Johansen2012,Pu2015}.
Indeed, it has been shown that the minimum spacing is mass dependent \citep{Weiss2018}, with a lower limit in observed Kepler systems of around 10 Hill radii.
As a result, understanding the mechanism leading to the instability of more tightly packed systems is critical to our understanding of planet formation and architecture.

The question of the stability of exoplanetary systems is particularly challenging due to several factors.
The observed close-in planets have most likely performed at least $10^{9}$ to $10^{11}$ orbits since their formation, which makes the numerical integration extremely costly if one wants to integrate the system over its whole lifetime.
Because of the age of exoplanetary systems, it is often assumed that the systems are stable to constrain the orbital configuration.
As a result, in order  to understand the architecture of planetary systems, the stability analysis is complementary to observations \citep{Laskar2017}.
Nevertheless, the process is made even more costly because we do not know the exact orbital configuration, let alone the planet masses for systems detected by transits.
But even if the present orbital configuration were known perfectly, planetary systems are chaotic, as has been shown for our own Solar System \citep{Laskar1994,Laskar2009}.
As a result, the only approach to a numerical stability analysis is to run several integrations with slight variations of the initial conditions to probe the outcome in a statistical manner.
Therefore, for each exoplanetary system, thousands of very costly numerical integrations would need to be run in order to obtain a satisfying understanding of its stability properties.
The process could eventually be sped up thanks to the help of machine learning classification \citep{Tamayo2016,Tamayo2020}.

Another approach is to rely on analytical stability criteria.
Under specific assumptions, it is possible to simplify the dynamics to obtain models accurately describing the behaviour of the system.
In particular, one can derive stability criteria that can delineate stable regions from unstable ones where systems will eventually experience close encounters and collisions.
Among such analytical criteria, one can cite the Hill stability \citep{Marchal1982,Gladman1993,Petit2018} and the overlap of mean motion resonances \citep[MMRs;][]{Wisdom1980,Deck2013,Petit2017,Hadden2018}.
For less compact, non-resonant systems, the dynamics are very well approximated by the secular model.
In the secular approximation, one averages over the fast motion of the planets on their Keplerian orbits to only consider their long-term deformations.
A well-known consequence of this averaging is the conservation of the planet semi-major axes, and thus of the angular momentum deficit \citep[AMD][]{Laskar1997,Laskar2000}.
The AMD gives a dynamically motivated measure of the total eccentricities and mutual inclinations in a planetary system, and thus acts as a dynamical temperature.
In particular, if the AMD is low enough, there is no possible orbital rearrangement allowing for close  planetary encounters.
This concept has been defined as the AMD-stability \citep{Laskar2017}; it allows for a fast characterisation of the stability of planetary systems away from MMRs, where the secular approximation is valid.
Besides the AMD-stability, the AMD has proven to be a versatile tool to understand planet dynamics \citep[\eg][]{Volk2020}.

However, the transition from the secular regime to regions where the fast interactions between planets shall not be neglected is unclear. 
This is due to the influence of  MMRs which forbid independent averaging over the fast angles of the planets  \citep[although it should be noted that an extension of the Lagrange-Laplace secular theory in the vicinity of MMRs is possible][]{Libert2013,Sansottera2019a}.
While theoretical studies in the two-planet case have allowed  a sharp limit to be found between the secular and non-secular regions \citep[][and references therein]{Hadden2018,Petit2018}, there are no complete studies for three-planet systems and higher.
Numerical simulations \citep[][and references in sec. \ref{sec:pheno}]{Chambers1996} have shown a qualitative change in behaviour between two-planet systems and three-planet systems and beyond: multi-planetary systems experience a long quiescent phase where the systems are almost secular before a very rapid transition to collisional dynamics.
Preliminary analytical studies were proposed by \cite{Zhou2007} and \cite{Quillen2011}, but their models did not entirely  reproduce the characteristics of the transition zone between long-lived systems and systems where scattering occurs immediately. 

The present work attempts to study the mechanism leading to the instability of tightly packed systems.
Since the different stability regime between two-planet and multi-planet systems starts at three planets, we focus on systems composed of three planets.
Contrary to previous studies, we do not make any assumptions regarding the masses of the planets (providing that they remain small) and consider unevenly spaced planets.
However, we restrict ourselves to initially circular and coplanar systems.
Indeed, due to interactions with the protoplanetary disk, compact, close-in systems most likely form in this state due to eccentricity and inclination damping \citep{Lin1986}.
We note that we do not consider planets trapped into resonant chains here and refer to \cite{Pichierri2020} for an analytical study of stability of resonant chains.
In addition, we are interested in systems that should be considered AMD-stable in the sense that no secular interactions can lead to their instability \citep{Petit2018}.
Understanding the initially circular systems gives a lower bound for the eccentric ones.
By analysing individual simulations, we postulate, as in \cite{Quillen2011}, that the instability is driven by the overlap of MMRs between the three planets of each system.
Their prominent role comes from the presence of a dense subset of three-planet resonances that covers a large part of the phase space, even for circular orbits.
Moreover, the system dynamics in the presence of this subset are not secular, yet they preserve the total AMD, which is a characteristic observed in numerical simulations. This feature is explained in Sect.\ \ref{sec:network-MMRHamiltonian}.
Using estimates of the diffusion rate proposed by \cite{Chirikov1979}, we are able to compute an analytical expression for the survival time.

Our analytical approach allows us to determine features in numerical simulations that trace the particular mechanism we study, which leads us to conclude that we isolated the right mechanism for planetary instability.
In particular, we confirm that the scaling in terms of Hill radius, widely used in numerical studies \citep{Chambers1996,Smith2009,Pu2015,Obertas2017}, is not appropriate.
By comparing with numerical simulations, we show that our time estimate is valid over four orders of magnitude in mass and almost seven orders of magnitude in survival time.

In the context of exoplanet observations, three-planet resonances are particularly significant as it is possible to assess their dynamical influence from transit data alone \citep{Delisle2017}. 
They can also be a signpost of the disruption of MMR chains thanks to tidal dissipation \citep{Charalambous2018,Pichierri2019}.
Nevertheless, the interactions between such resonances has not been exhaustively studied. 

In section \ref{sec:pheno}, we begin by a review of the works on the problem of tightly packed planetary systems and we perform an in-depth qualitative analysis of the instability.
In section \ref{sec:technicalsetup}, we introduce our framework to treat the problem of three-planet MMR.
Section \ref{sec:network} contains most of the technical details. We first describe the network of zeroth-order three-planet resonances. We then solve the dynamics for an isolated  MMR to finally obtain a criterion delimiting the region where the MMRs overlap.
Using the framework developed in section \ref{sec:network}, we estimate in section \ref{sec:timescale} the survival time for a system of three planets, with arbitrary mass distribution and spacing (assuming that the planets are not too massive and tightly packed).
We compare our analytical results to numerical simulations in section \ref{sec:EMScomp}.
Finally, we discuss possible extensions to more general systems than three planets on circular and coplanar orbits in section \ref{sec:beyond}.
While the analytical derivations make it necessary to define auxiliary variables, we tried where possible to use only variables with a clear physical meaning in the figures to help those readers willing to skip the technical sections.

\section{Qualitative description of the instability}
\label{sec:pheno}

The dynamics of tightly packed systems are chaotic, and research on the subject has mainly focused on a qualitative description of their behaviour due to the difficulty of the analytic approach.
We review the qualitative description proposed by previous studies and highlight how the instability is triggered.

\subsection{Stability in the two-planet case}

While the three-body problem is not integrable in general, the problem of the stability of a two planet system is well understood.
Most of the stability results come from the existence of a topological boundary in the three-body configuration space leading to the so-called Hill-stability \citep{Marchal1982}.
In a Hill-stable system, the two planets can never approach one another, which leads to a sharp difference in behaviour.
The Hill-stability was popularised by \cite{Gladman1993} for circular orbits, as a minimal distance between orbits normalised by their Hill radius guaranteeing the system's stability. This stability criterion can be written as
\begin{equation}
\Delta = \frac{a_2-a_1 }{a_1} > 2\sqrt{3} \frac{R_H}{a_1} \simeq 3.46 \frac{R_H}{a_1},
\label{eq:Hill}
\end{equation}
where
\begin{equation}
R_H = \frac{a_1+a_2}{2}\left(\frac{m_1+m_2}{3m_0}\right)^{1/3} 
\label{eq:Hillrad}
\end{equation}
is the mutual Hill radius with $m_1,m_2$ being the planet masses and $m_0$, the star mass.
For inclined and eccentric orbits, there exists a critical AMD value depending only on semi-major axis and masses such that a system with a smaller AMD is Hill stable \citep{Petit2018}.

Another stability criterion for a two-planet system can be derived from the overlap of MMR \citep{Wisdom1980,Deck2013,Petit2017,Hadden2018}. 
While the unperturbed resonant problem is integrable, the interaction between neighbouring MMRs leads to the formation of a chaotic web such that the planets' orbital elements wander in a random walk fashion.
This behaviour is known as the \cite{Chirikov1979} diffusion.
For initially circular orbits, the overlap occurs at a distance scaling as $\left((m_1+m_2)/m_0\right)^{2/7}$ \citep{Wisdom1980}.
The exponent 2/7 is close to 1/3 but it has been highlighted that there exists a regime where MMRs overlap while the planets are Hill-stable, that is,\ the system is long-lived while experiencing short-term chaos \citep{Deck2013,Petit2018}.

This means that a two-planet system is either stable over timescales comparable with the lifetime of the host star or unstable in a very short amount of time (less than $10^5$ revolutions). No such dichotomy is observed for multiplanetary systems.
Indeed, a multiplanet system can appear stable if it is numerically integrated over a few million orbits while becoming unstable in less than a billion years.

\subsection{Survival time of tightly packed systems}

The pioneering work on the stability of tightly packed multi-planetary systems was carried out by \cite{Chambers1996}.
These latter authors performed numerical simulations of systems with equal-mass planets on initially equally spaced circular and coplanar orbits (hereafter referred to as EMS systems).
The constant orbit spacing is given by $\Delta = (a_{k+1}-a_k)/a_k$.
For various planetary masses and numbers of planets, these latter authors recorded the survival time of a system, defined as the integration time before the distance between the two planets becomes smaller than a Hill radius.
As shown by \cite{Rice2018}, the time between such a close encounter and the proper collision is usually negligible. 
\cite{Chambers1996} observed that the survival time grows exponentially with the spacing $\Delta$ rescaled by the Hill radius $R_H$,
\begin{equation}
\log_{10} \frac{\Tsurv}{P} = b\frac{\Delta}{R_H} + c
\label{eq:Tsurv_hill_exp}
,\end{equation}
where $P$ is a typical orbital period and $b$ and $c$ are numerical factors\footnote{Throughout this paper, $\log_{10}$ designates the decimal logarithm and $\ln$ the natural logarithm  in base $\EXP$.}. Here, $b$ seems to have a small dependency on the mass ratio and the number of planets, and
$c$  also seems to depend on the mass ratio.
Analysing Fig. 4 from \citep{Chambers1996}, a more appropriate scaling seems to be
\begin{equation}
\log_{10} \frac{\Tsurv}{P} = b'\Delta \left(\frac{m_p}{m_0}\right)^{-1/4} - c' - \log\frac{m_p}{m_0},
\label{eq:Tsurv_good_exp}
\end{equation}
where $b'$ and $c'$ are positive numerical coefficients independent of the masses, $m_p$ is the planet mass and $m_0$ the star mass. We note that such scaling was also chosen by \cite{Faber2007}.

Subsequent numerical works on the stability of EMS have been carried out.
As the computational capacities increased, \cite{Smith2009} and then \cite{Obertas2017} obtained datasets with a much finer distribution of spacing and longer integration times showing systems becoming unstable after almost 10 Gyr.
Beyond the trend already observed by \cite{Chambers1996}, \cite{Obertas2017} showed that the survival time is reduced in the vicinity of low-order two-planet MMR. 
\cite{Hussain2020} show that the spreading around the linear trend for $\log\Tsurv/P$ is roughly constant and follows a normal distribution with a standard deviation of $0.43\pm 0.16$~dex, indicating that the instability emerges from a chaotic diffusion process.
Beyond the EMS initial conditions, \cite{Pu2015} explored the impact of small variations of the initial conditions by drawing the spacings, eccentricities, and inclinations from distributions and showed that the exact spacing can be replaced by the minimal separation between the orbits.
From these studies, the minimal spacing ensuring the stability over a few billion orbits can be estimated to be around 10 Hill radii.

Following \cite{Chambers1996}, most of the previously cited studies fit the survival time with curves similar to Eq. \eqref{eq:Tsurv_hill_exp} because of the natural parallel with the two-planet case.
However, there is no generalisation of the Hill stability in the multi-planet case and the mechanism leading to instability has a priori no reason to be related to the Hill scaling $R_H$.
The discrepancy between the two proposed mass renormalisations in Eqs. \eqref{eq:Tsurv_hill_exp} and \eqref{eq:Tsurv_good_exp} is easily explained by the fact that most studies only considered a limited mass range and very small difference between the exponents.
\cite{Zhou2007} estimated \Tsurv\ as a power-law in the spacing and using Nekhoroshev estimates, \cite{Yalinewich2020} proposed a scaling similar to Eq. \eqref{eq:Tsurv_good_exp}.

\subsection{Phenomenology of the instability}

These qualitative and quantitative studies on EMS systems highlight the key features that the tightly packed system instability presents and that an analytical model should explain.

\begin{enumerate}[a.]
        \item The survival time $\Tsurv$ seems to have an exponential dependency on the orbital spacing, measured in units of $(m_p/m_0)^{1/4}$. 
        The fit is valid over 6 to 8 orders of magnitude for survival times between 100 and almost $10^{10}$ orbits. The higher end is limited by computational time. However, the physical interest to go beyond is limited as it approaches the lifetime of the central star in most cases.
	
        \item Instabilities occur for spacings larger than that leading to two-planet instabilities. As a result, it is an intrinsically multi-planet phenomenon.
        In addition, \cite{Chambers1996} have shown that the results were unchanged in systems of four or more planets if the planet interactions are limited to their neighbours.
        Thus, three planets are necessary but also sufficient to reproduce the effect.
	
        \item Systems initially on circular orbits, and therefore AMD-stable, can become unstable. The mechanism at play is thus by nature non-secular and involves some kind of MMR overlap despite the fact that two-planet MMRs do not overlap in the range where the instability can occur. However, the AMD does not evolve regularly during the lifetime of the system. Indeed, as shown in Figure \ref{fig:pheno-periodrat}, a system can experience almost no AMD evolution during most of its lifetime before a rapid increase shortly before instability.
	
        \item The survival time distribution suggests that the evolution is driven by a diffusion process \citep{Hussain2020}. The dips close to first-order two-planet MMRs indicate that these latter play a fundamental role in enabling the orbit crossing.
\end{enumerate}

While the stability of EMS systems has been described extensively from numerical simulations, very few works have developed an analytical framework attempting to describe the observed behaviour.
In the most elaborate model, \cite{Quillen2011} proposed that the instability is driven by the overlap of zeroth-order three-planet MMRs.
Resonances involving more than two planets emerge as the result of the first-order averaging \citep[\eg\ Chapter 2,][see also section \ref{sec:technicalsetup}]{Morbidelli2002} and are weaker than the two-planet MMR.
\cite{Quillen2011} shows that, despite their smaller width, the three-planet MMRs are more numerous and overlap at larger spacing and smaller eccentricities.
The ansatz is that the semi-major axes of the planets evolve randomly through the rich network of these three-planet MMRs until a first-order two-planet MMR is encountered, leading to a rapid AMD increase, and  close encounters and collision shortly afterwards.
Moreover, the main resonances close to circular orbit preserve the total AMD (see section \ref{sec:network}), which is consistent with simulations.
We highlight the fact that, in the secular dynamics of the Solar System, slow diffusion leads to a region where the system becomes rapidly unstable  \citep{Laskar1994,Batygin2015a}.

To illustrate the mechanism leading to instability, we perform the numerical integration of a typical EMS system.
The planets have a mass $m_p=10^{-5}\ \Ms$ and orbit a solar-mass star. 
The inner orbit is at 1 au and the period ratios between adjacent planets is initially close\footnote{The initial period ratios are not rigorously equal in this specific example.} to  $P_{k+1}/P_k = 1.175$.
This particular value was chosen in order to observe the instability after roughly a few million orbits of the inner planet while being outside of a two-planet MMR island.
The orbits are initially circular and coplanar and the angles drawn randomly.
As in previous studies, we run the simulations up to the first close encounter. In the considered case, the integration lasts 3.33 Myr.
The system is integrated with the hybrid integrator \texttt{MERCURIUS} \citep{Rein2019a} from the \texttt{REBOUND} code \citep{Rein2012a} with a time-step of 0.01 yr.
The relative energy error is $5\times10^{-10}$.

\begin{figure}
	\includegraphics[width=\linewidth]{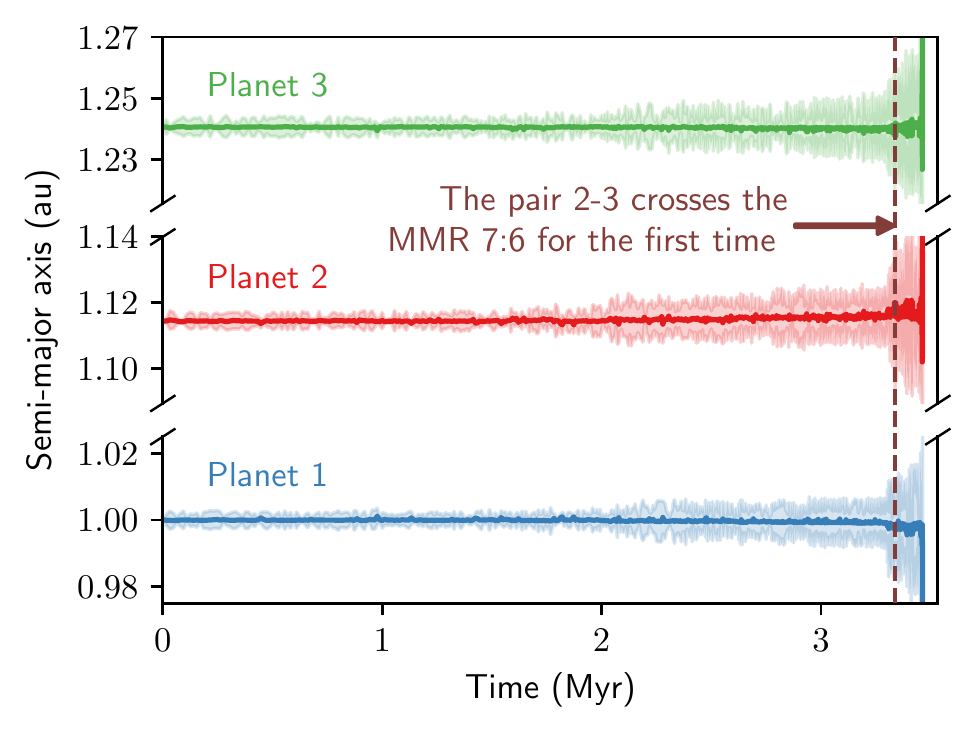}
        \caption{Semi-major axis as a function of time for an example of a three-planet EMS system (see the text for the full initial conditions). The envelope of the curve represents the extent of the orbits. We note the discontinuous vertical axis. We show the time where the first main two-planet MMR is crossed. The system becomes unstable soon afterwards.\label{fig:pheno-sma}}
\end{figure}

Figure \ref{fig:pheno-sma} shows the evolution of the semi-major axis of the three planets.
The envelope around the curve corresponds to the extent of the orbits, that is,\ the position of the periapses and apoapses, and is thus a measure of the eccentricities of the orbits.
The curves are smoothed by performing a rolling averaging over the next ten snapshots\footnote{This method is a proxy to highlight the evolution of the averaged canonical coordinates defined in section \ref{sec:technicalsetup}, as it removes the rapid oscillations due to the fast dynamics onto the orbits. This is sufficient here as our goal is to illustrate the mechanism at play.}.
The vertical axis is discontinuous to highlight the small variations during the large majority of the integration.
As already described by previous authors, the system appears quiescent for the majority of its lifetime.
Subsequently, after the pair 2-3 crosses the 7:6 resonance, the system becomes unstable in 127 kyr.
This figure emphasises the timescale difference between the lifetime of the system and the proper unstable phase that is almost two orders of magnitude shorter.
Explaining the lifetime of tightly packed systems should therefore focus on the quiescent phase as the timescale to reach instability is dominated by this phase.
\begin{figure}
	\includegraphics[width=\linewidth]{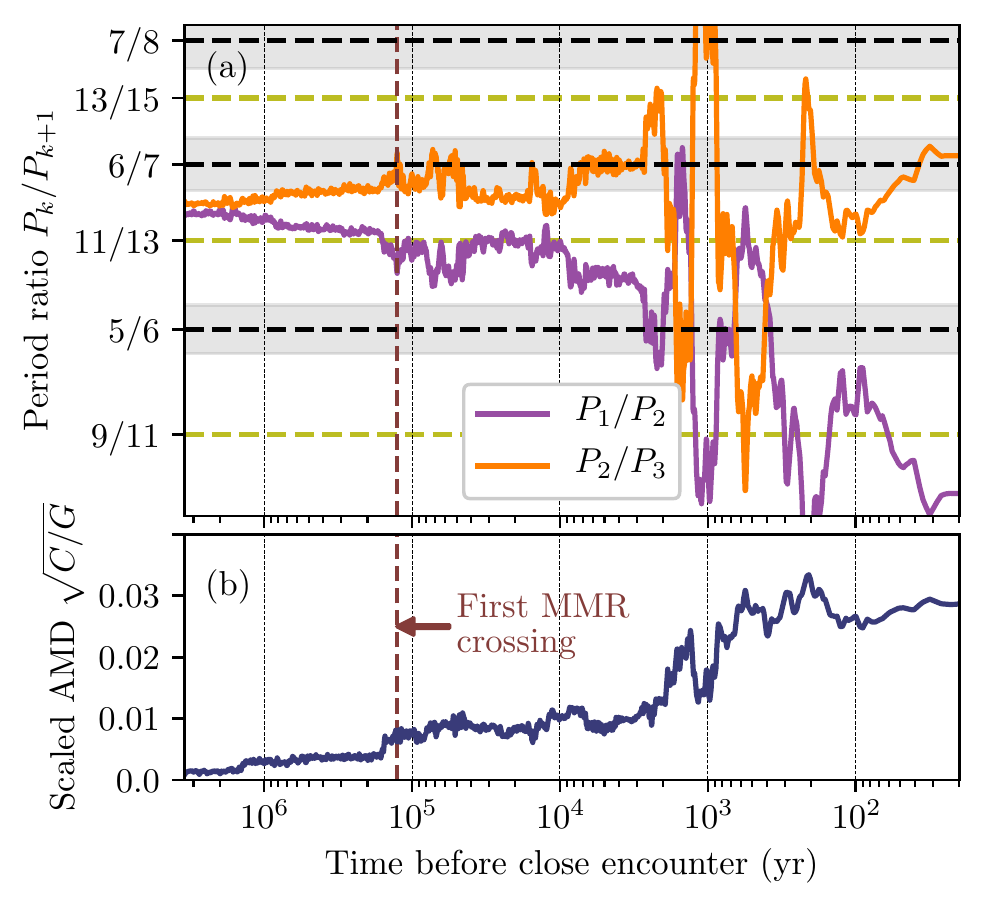}
        \caption{Panel (a) Period ratio of the adjacent pairs as a function of the time to the close encounter.  We note the logarithmic scale. The vertical dashed line is the same as in figure \ref{fig:pheno-sma}. The black horizontal dashed lines corresponds to first-order MMRs, the yellow dashed lines to the second-order MMRs.
                The width of the first-order MMRs is displayed in grey.
                Panel (b) AMD normalised by the total angular momentum as a function of the time to the close encounter. Using the square root gives a typical value of the planet eccentricities.\label{fig:pheno-periodrat}}
\end{figure}

To show the rapid change of behaviour before the close encounter, Figure \ref{fig:pheno-periodrat}a shows the evolution of the two adjacent period ratios $P_k/P_{k+1}$ as a function of the time to the close encounter (we note the logarithmic scale).
In Figure \ref{fig:pheno-periodrat}b, we plot the evolution of the  AMD $\AMD$ of the system (see eq. \ref{eq:AM-AMD}) rescaled by the total angular momentum $\AM$. The plotted quantity, $\sqrt{\AMD/\AM}$ , scales linearly with eccentricity for close to circular orbits.
At the moment when the pair 2-3 enters the 7:6 MMR region, the system enters the scattering phase. This is also the moment where the AMD starts to increase.
Nevertheless, the initial phase is not secular despite the near conservation of the AMD; indeed we see that the period ratios are not constant but evolve over a long timescale. 

\begin{figure}
	\includegraphics[width=\linewidth]{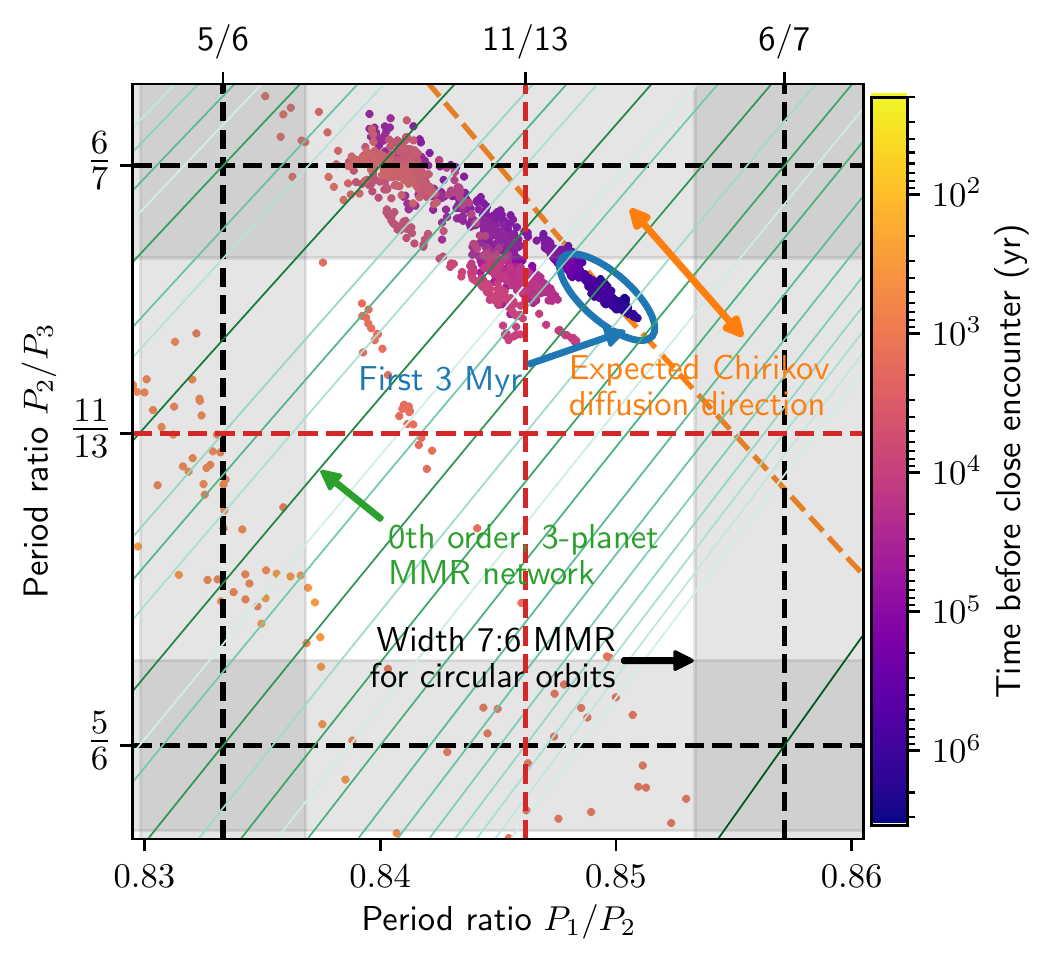}
        \caption{Evolution in the period ratios plane. The points are colour-coded according to their time before the close encounter.
                We note that the system spends almost all of its time very close to its starting location as shown in Figs. \ref{fig:pheno-sma} and \ref{fig:pheno-periodrat}.
                 Green oblique lines correspond the loci of the zeroth-order three-planet MMRs (see section \ref{sec:network-descr}).
                 Chirikov's diffusion is expected to occur perpendicularly to the network direction, along the dashed orange line.
        The extent of the adjacent first-order two-planet  MMRs, 7:6 and 6:5, is plotted in grey. The width is computed for circular orbits. The second-order resonance 13:11 is plotted in red. 
        Once the system enters the two-planet resonance network, the diffusion is much more rapid.\label{fig:pheno-network}}
\end{figure}

The interaction and the position of the system with respect to the network of two-planet MMRs seems critical to the duration of the quiescent phase.
However, Fig. \ref{fig:pheno-periodrat} merely shows how the instability is triggered and not the mechanism leading to it.
The slow evolution of the system is seen much more explicitly in the period ratio plane plotted in Figure \ref{fig:pheno-network}.
We plot the period ratio of the outer pair $P_2/P_3$ as a function of the inner pair period ratio $P_1/P_2$.
In this plane, the two planet MMRs are vertical and horizontal dashed lines. We plot the neighbouring first-order MMR, that is,\ the 7:6 and the 6:5 in black, indicating their approximate extent for the circular orbit in grey.
The second-order resonance 13:11 is plotted in red, but it has a null width for circular orbits.
We see the system starts outside of the two-planet MMR.
However, on top of the two-planet MMR network, there also exists the network of three-planet MMRs. For circular orbits, the main three-planet MMRs are of zeroth order (see section \ref{sec:network-descr}).
We plot the loci of the largest three-planet MMR in the vicinity of the initial condition: this network is composed of a set of almost parallel lines which run transversally to the two-planet MMR lines.
As predicted by \cite{Chirikov1979} theory, the diffusion takes place perpendicularly to the network of the three-planet MMRs, up until the system reaches the two-planet resonances where the trajectory wanders around rapidly.

This qualitative analysis seems to confirm Quillen's hypothesis. 
The survival time is dominated by the diffusion along the three-planet MMR network and the system becomes unstable once it reaches the two-planet resonance where chaotic diffusion is faster and  rapidly increases the total AMD.
The survival time can be estimated by computing the diffusion rate according to Chirikov's resonance overlap theory (see section \ref{sec:diffusion-chirikov}).
The scaling law  for the survival time obtained by \cite{Quillen2011} is a very steep power-law instead of having an exponential behaviour.
In particular, the timescale is overestimated at short separations and underestimated for large ones.
Quillen's result and its difference with numerical simulations can be explained by some simplifications made in the computations leading to an inexact determination of the effective diffusion rate as well as a limit of the three-planet MMR overlap.

In this study, we consider the general, circular, coplanar three-planet problem. We remain in the framework of tightly packed systems but we relax the assumption on the initial equal spacing and equal masses.
We show that it is possible to use Chirikov's theory to explain the observed survival time scaling.

\section{Problem considered and mean motion resonances}
\label{sec:technicalsetup}

We summarise most of the notations in Table \ref{tab:notations}.
We consider a system of three planets of masses $m_1,m_2$ , and $m_3$ orbiting a star of mass $m_0$.
The canonical positions $\vec{r}_j$ and momenta $\vec{\tilde{r}}_j$ are expressed in canonical heliocentric coordinates \citep{Poincare1905,Laskar1991}.
The initial orbits are assumed to be circular and coplanar. 
Let the semi-major axes $a_j$, the eccentricities $e_j$, the mean longitudes $\lambda_j$  , and the periapses longitude $\varpi_j$ be the orbital elements defining the orbits.
A set of canonical coordinates for the system is given by the modified Delaunay coordinates \citep[\eg][]{Laskar1991}:
\begin{align}
\Lambda_j &= m_j\sqrt{\mu a_j},  &\lambda_j,\nnb
\AMD_j & = \Lambda_j\left(1-\sqrt{1-e_j^2}\right), &-\varpi_j,\label{eq:Delaunay}
\end{align}
where $\mu = \Gr m_0$ and $\Gr$ is the gravitational constant.
We note that the gravitational parameter $\mu$ is the same for all three planets as in \cite{Laskar2017}.
This is possible if we consider the so-called democratic-heliocentric formulation of the planetary Hamiltonian \citep[\eg][]{Morbidelli2002}.
The couples of variables $(\AMD_j,-\varpi_j)$ can also be replaced by their associated complex variables 
\begin{align}
x_j &= \sqrt{\AMD_j}\EXP^{\i \varpi_j},& -\i\bar{x}_j,
\label{eq:Poincare}
\end{align}
with $\i=\sqrt{-1}$ ($x_j$ are the canonical momenta and $-\i\bar{x}_j$ the conjugated positions). For small eccentricities, we have ${x_j\simeq \sqrt{\Lambda_j/2}e_j\EXP^{\i \varpi_j}}$. The system total angular momentum $\AM$ and AMD $C$ are given by
\begin{equation}
\AM = \sum_{j=1}^{3}(\Lambda_j-\AMD_j) \quad \text{and}\quad \AMD= \sum_{j=1}^{3}\AMD_j.
\label{eq:AM-AMD}
\end{equation}

The Hamiltonian $\H$ describing the dynamics can be split into an integrable part  
\begin{equation}
\H_0 = \sum_{j=1}^{3}\frac{\|\vec{\tilde{r}}_j\|^2}{2m_j}-\frac{\mu m_j}{r_j} = -\sum_{j=1}^3 \frac{\mu^2m_j^3}{2\Lambda_j}
\label{eq:HKeplerian}
,\end{equation} 
describing the motion on unperturbed Keplerian orbits, and a perturbation, \begin{equation}
\epsilon \H_1  = -\sum_{j<j'}\frac{\Gr m_jm_{j'}}{|\vec{r}_j-\vec{r}_{j'}|}  + \frac{1}{2m_0}\left\|\sum_{j=1}^3\vec{\tilde{r}}_j\right\|^2
\label{eq:Hpert}
,\end{equation}
describing the planet interactions. Here, $\epsilon$ is a dimensionless parameter of the order of the planet-to star-mass-ratio to reflect the scale difference between the two parts of the Hamiltonian.
In terms of Poincar\'e coordinates, the perturbation part can be written as
\begin{equation}
\epsilon\H_1 =  \sum_{\mathbf{k},\mathbf{l},\mathbf{\bar l},} \mathcal{C}_{\mathbf{k},\mathbf{l},\mathbf{\bar l}}(m_j,\Lambda_j) \left(\prod_{j=1}^3 x_j^{l_j}\bar{x}_j^{\bar l_j}\right)\EXP^{\i \mathbf{k}\cdot\boldsymbol{\lambda}},
\label{eq:H1.Poin}
\end{equation}
where $\mathbf{k}\in \Z^3$, $\mathbf{l},\mathbf{\bar l}\in \N^3$.

Due to the conservation of angular momentum, the coefficient $\mathcal{C}_{\mathbf{k},\mathbf{l},\mathbf{\bar l}}$ must vanish unless the indices $\mathbf{k}, \mathbf{l},\mathbf{\bar l}$ verify the d'Alembert rules \citep[\eg][]{Morbidelli2002} and in particular
\begin{equation}
\sum_{j=1}^{3} k_j +l_j-\bar{l}_j = 0.
\label{eq:dalembert}
\end{equation}

In the unperturbed case, the system is said to be in a MMR if the mean motions 
\begin{equation}
n_j =\dot{\lambda}_j =  \dpart{\H_0}{\Lambda_j} = \frac{\mu^2m_j^3}{\Lambda_j^3}
\label{eq:meanmotion}
\end{equation}
verify an equation of the form
\begin{equation}
\sum_{j=1}^3 k_jn_j = 0.
\label{eq:reseq}
\end{equation}
The sum $k=k_1+k_2+k_3$ is the {`order'} of the resonance.
The sum $K = |k_1| + |k_2| + |k_3|$ is the {`index'} of the resonance.
In the general case, the perturbation $\epsilon\H_1$ also influences the resonant dynamics.
The terms in eq. \eqref{eq:H1.Poin} contributing to the resonance are the ones that depend on the combination of mean longitudes $k_1\lambda_1+k_2\lambda_2+k_3\lambda_3$.
Because of d'Alembert rules, the leading order term in the perturbation is of order $k$ in eccentricities ($k$ being the resonance order).

We note that because each term in  eq. \eqref{eq:Hpert} only contains contributions from two planets, this is also the case for \eqref{eq:H1.Poin}.
In particular, there are no terms in  the non-averaged Hamiltonian $\H=\H_0+\epsilon\H_1$ that depend on angles of the form $\mathbf{k\cdot\boldsymbol\lambda}$ with $k_j\neq 0$ for all $j$.
This means that there are no three-planet resonances at the first order in $\epsilon$. There are instead of course $\mathcal O(\epsilon)$ two-planet MMR terms.

Three-planet MMRs actually emerge in the perturbative Hamiltonian as $\mathcal O(\epsilon^2)$ terms which appear after applying a perturbation step which eliminates the fast angles $\boldsymbol\lambda$ to first order in $\epsilon$ (this step is sometimes referred to as \emph{averaging} because to first order in $\epsilon$ it is equivalent to averaging out the fast angles from the Hamiltonian).
To do so, we assume that  the system is `far enough' from the first-order two-planet MMR\footnote{We simplify the general case  because the first-order MMRs are the only ones with a non-zero width for close to circular orbits. In general, we should require that $k_jn_j+k_{j'}n_{j'}$ is small with respect to the largest coefficient corresponding to this particular order in the sum eq. \eqref{eq:H1.Poin}. But because of their dependence on eccentricity, such a coefficient is negligible for higher order resonances. } , that is,\ we assume that $kn_j-(k+1)n_{j'}$ is not too small with respect to $\epsilon$ for some integer $k$ \citep{Morbidelli2002}.
This is for example the case for the system considered in the previous section. We sketch the main lines of these perturbative steps below, and we carry out the explicit calculation of the relevant terms in the following section.

As we consider systems far enough from two-planet resonances, we can perform one perturbation step and keep track of all terms up to order $\mathcal O(\epsilon^2)$ in the equations.
We use the classical approach from the perturbations theory, the Lie series method \citep{Deprit1969}. We refer to section 2.2 of \cite{Morbidelli2002}  and references therein for a complete description of the method. The method has already been applied to provide an analytical model of three-body resonances when one of the bodies is a test particle \citep{Nesvorny1998}.
The idea is to introduce a new set of variables (noted with a prime in the following equations) $\epsilon$-close to the original ones such that in these new variables the transformed Hamiltonian writes
\begin{equation}
\H' = H_0(\boldsymbol\Lambda') +\epsilon\bar{\H}_1(\boldsymbol\Lambda',\mathbf x') +\epsilon^2\H_2'(\boldsymbol\Lambda',\boldsymbol\lambda',\mathbf x') +\O(\epsilon^3),
\end{equation}
where $\epsilon\bar{\H}_1$ is the average of $\epsilon\H_1$ over the mean longitudes $\boldsymbol\lambda$, and $\epsilon^2\H_2'$ is the leading-order term of a series in $\epsilon$.
The transformation can be explicitly constructed as the flow between 0 and 1 of a generating Hamiltonian vector field $\exp(\{\epsilon\chi_1,\cdot\})$ where $\{\cdot,\cdot\}$ is the Poisson bracket\footnote{We use the convention $\{f,g\} = \sum_j \left(\dpart{f}{p_j}\dpart{g}{q_j}- \dpart{f}{q_j}\dpart{g}{p_j}\right)$ where $(\vec{p},\vec{q})$ is a set of conjugated coordinates.} and $\epsilon\chi_1$ is the solution of the homological equation
\begin{equation}
\left\{\epsilon \chi_1,\H_0\right\} +\epsilon \H_1 = \epsilon\bar{\H}_1.
\label{eq:homological}
\end{equation}
More precisely, if we note $\hk=\hk(\mathbf x,\bar{\mathbf x}),$ the complex Fourier coefficients of $\H_1$ with respect to the mean longitudes,
we can write
\begin{equation}
\epsilon\chi_1 = \epsilon\sum_{\vec{k}\neq0} \frac{\hk}{\i\vec{k}\cdot\vec{n}}\EXP^{\i\vec{k}\cdot\boldsymbol\lambda}.
\label{eq:chi}
\end{equation}
Due to the expression of $\epsilon\H_1$ given in eq. \eqref{eq:H1.Poin}, the denominators $\vec{k}\cdot\vec{n}$ are of the form $k_jn_j+k_{j'}n_{j'}$ and are not `too small' because we assume the system to be far from two-planet MMRs. Thus the formal series \eqref{eq:chi} is formally well defined; one can stop the summation at indices $\vec{k}$ of sufficiently high order so that the remaining Fourier terms in $\H_1$ have sizes smaller than $\epsilon^2$, which is ensured by the exponential decay of the Fourier coefficients. Thus, the solution \eqref{eq:chi} to the homological equation \eqref{eq:homological} is well defined.
We can express the  Hamiltonian $\epsilon^2\H_2'$ as
\begin{equation}
\epsilon^2 \H_2' =\frac{1}{2}\left\{\epsilon \chi_1, \epsilon \H_1 + \epsilon \bar \H_1\right\}.
\label{eq:H2full}
\end{equation}
The Poisson bracket in eq. \eqref{eq:H2full} generates terms involving all three mean longitudes. In other words, three-planet MMRs that were not present in the initial Hamiltonian cannot be neglected at second order in averaging.
The study of a particular three-planet MMR can be done by a second averaging over the other fast angles, because all other terms will not contribute small divisors and can therefore be eliminated by an additional perturbative step. 
In practice, this results in another change of coordinates, which are $\epsilon^2$ close to the first-order averaged coordinates, and the new Hamiltonian is the average of $\H'$ with respect to the fast (\eg\ non resonant) angles (see following section).
Because we do not need to change back to the initial coordinates,  hereafter we drop the primes on the coordinates and Hamiltonian.
We also drop the terms of order $\epsilon^3$ and greater.

\section{The three-planet zeroth-order resonance network}
\label{sec:network}

In figure \ref{fig:pheno-network}, we see that the diffusion mainly occurs perpendicularly to the zeroth-order three-planet MMRs.
This is expected for close to circular orbits because the resonant coefficients do not depend on eccentricity at the leading order in eccentricity.
In addition, the structure of the network is easier to describe.
We make the hypothesis that the zeroth-order three-planet MMRs are sufficient to explain most of the diffusion leading to the instability. This assumption is well supported in section \ref{sec:EMScomp}, where we compare the analytical prediction of survival times calculated under this hypothesis with the results of numerical simulations.
We analyse these MMRs and compute an overlap criterion in this section.
We consider the role of additional MMRs in section \ref{sec:beyond}.

\subsection{Network description}
\label{sec:network-descr}

A zeroth-order three-planet MMR can be described by two integers $p$ and $q$. The resonance equation is
\begin{equation}
pn_1-(p+q)n_2+qn_3 = 0.
\label{eq:03pMMR}
\end{equation}
Since such resonance does not depend on the longitude of the periapses, the AMD is unaffected by the resonant terms (see below).
We therefore restrict ourselves to the three\footnote{Each couple of conjugated variables counts for one degree of freedom.} degrees of freedom $(\Lambda_j,\lambda_j)$.
The resonance equation defines a plane in the frequency space $(n_1,n_2,n_3)$.
Because the gravitational interactions are scale invariant, we can restrict ourselves to a two-dimensional plane corresponding to the period ratios \perrat{12} and \perrat{23} where 
\begin{equation}
\perrat{ij} = \frac{P_i}{P_j}= \frac{n_j}{n_i}.
\label{eq:perrat}
\end{equation}
Dividing eq. \eqref{eq:03pMMR} by $n_2$, and reorganising terms, one gets
\begin{equation}
\perrat{23} = 1-\frac{p}{q}(\perrat{12}^{-1}-1),
\label{eq:03plMMRperrat}
\end{equation}
that is the equation of a straight line passing through the point (1,1)  with slope $-p/q$ for the period ratios $\perrat{21} = P_2/P_1>1$ and $\perrat{23} = P_2/P_3<1$ .
While the resonances can be interpreted easily geometrically with these two period ratios, the fact that one of the variables is larger than 1 and the other smaller can be confusing.
Moreover, expanding the period ratio \perrat{21}  for tightly packed planets (\ie\ close to 1) leads to a poorer approximation at first order than expanding  \perrat{12}.
We therefore only consider the variables \perrat{12} and \perrat{23} as done in eq. \eqref{eq:03plMMRperrat}.
In the plane (\perrat{12},\perrat{23}), the resonance loci are hyperbolas passing through (1,1); they however behave to a very good approximation as straight lines with slopes $p/q$ for tightly packed systems.
As shown in the following section, the strength of the resonances depends strongly on their index $2(p+q)$.

\begin{figure}
	\includegraphics[width=\linewidth]{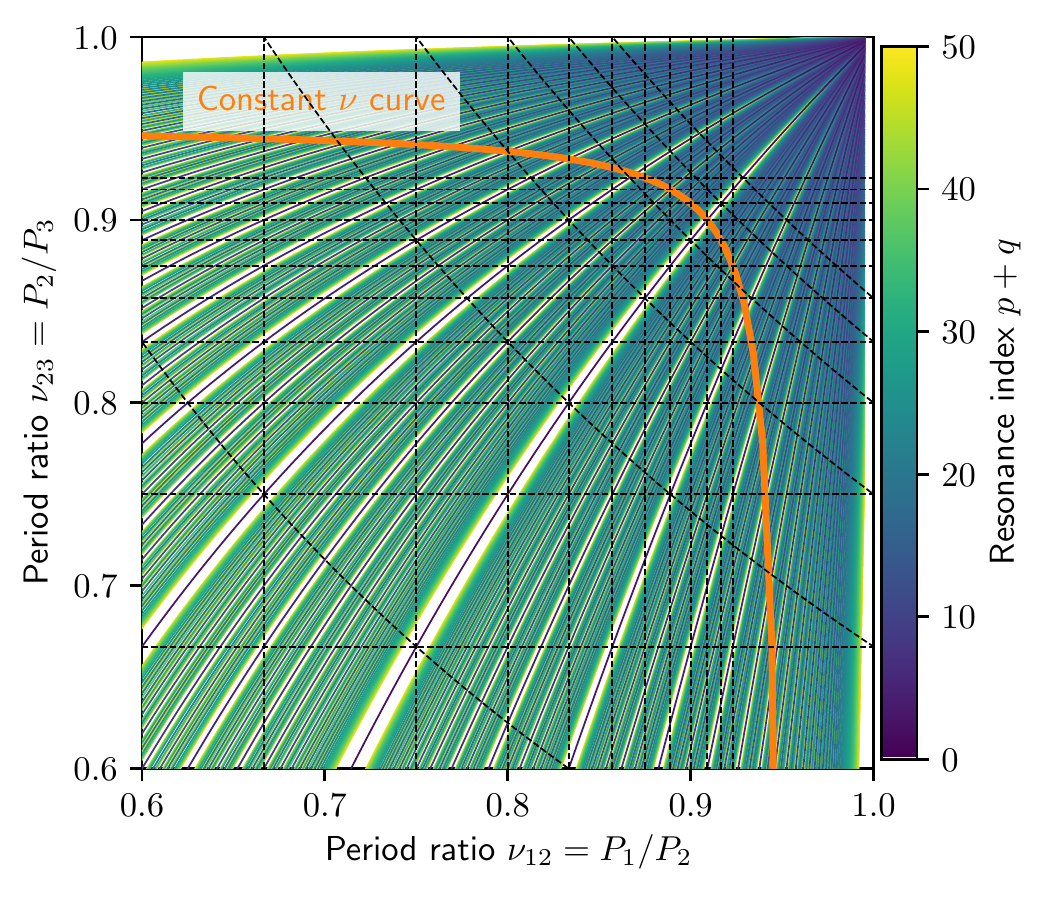}
        \caption{Zeroth-order three-planet MMR loci in the period ratio plane. The colour indicates the index $p+q$ of the resonance. The dashed lines corresponds to first-order two-planet MMRs (the oblique ones correspond to MMRs between planets 1 and 3).
	The curve  $\genperrat=0.05$ is displayed in orange (see eq. \ref{eq:genperrat}).
		\label{fig:network}}
\end{figure}

Figure \ref{fig:network} shows the loci of the zeroth-order three-planet MMRs such that  $p+q<50$ (the dashed black lines correspond to two-planet first-order MMR). Because the slope at the point $(1,1)$ is $p/q$, the resonances are not spread uniformly. Indeed, higher index resonances can lie on top of lower index ones if $p$ and $q$ are not coprime, for example where\ the resonance $2n_1-4n_2+2n_3=0$ is the same as the resonance $n_1-2n_2+n_3=0$.
The full network  is dense in the $(\perrat{12},\perrat{23})$ plane. 
However, the resonance width may become so small as $p+q$ increases that the resonances do not interact. 

The resonance loci do not depend on the MMR index, but only on the ratio $p/(p+q)$.
We choose this specific ratio as it lies between 0 and 1.
One can see that  $p/(p+q)$ can be extended as a continuous function in the period ratio plane.
An adapted set of coordinates to describe the period ratio plane can be defined\footnote{These coordinates are not canonical but are nevertheless convenient to describe the period ratio plane.} to take advantage of this property.
We define the resonance locator
\begin{align}
\resloc = \frac{1-\perrat{23}}{\perrat{12}^{-1}-\perrat{23}} = \frac{\perrat{12}(1-\perrat{23})}{1-\perrat{12}\perrat{23}}= \frac{p}{p+q}.
\label{eq:resloc}
\end{align}
The second equality is only valid on resonances. The position along the resonance can be defined by a generalised period ratio separation
\begin{equation}
\genperrat = \frac{(1-\perrat{12})(1-\perrat{23})}{1-\perrat{12}\perrat{23}} = \frac{p}{p+q}(\perrat{12}^{-1}-1) = \frac{q}{p+q}(1-\perrat{23}),
\label{eq:genperrat}
\end{equation}
where $\resloc$ is a constant on a specific resonance whereas $\genperrat$ is a hyperbola along which the resonance strength is roughly comparable.

The variables $(\genperrat,\resloc)$ are well adapted to describe the dynamics governed by the three-planet MMRs.
We can express the period ratios as a function of these variables
\begin{equation}
\perrat{12} = \frac{\resloc}{\resloc+\genperrat}, \quad \text{and} \quad \perrat{23} = \frac{1-\resloc-\genperrat}{1-\resloc}.
\label{eq:etanutoperrat}
\end{equation}
The levels of constant $\genperrat$ are hyperbola with horizontal and vertical asymptotes $(1+\genperrat)^{-1}$. The curve $\genperrat=0.05$ is 
shown in Fig. \ref{fig:network}  in orange.

\subsection{Single zeroth-order three-planet MMR Hamiltonian}
\label{sec:network-MMRHamiltonian}
Let us consider a specific resonance described by the integers $p$ and $q$.
One can make a linear change of variables to use explicitly the resonant angle and average over the non-resonant ones.
Let us define
\begin{align}
\theres &=   p \lambda_1 - (p+q)\lambda_2 +q\lambda_3,\nnb
\theta_\scalefactor &= \lambda_2 - \lambda_3,\label{eq:transf-angles}\\
\theta_\CAM & = \lambda_3.\nonumber
\end{align}
The conjugated momenta are
\begin{align}
\resact &=   \frac{1}{p}\Lambda_1,\nnb
\scalefactor &=\frac{p+q}{p} \Lambda_1 + \Lambda_2,\label{eq:transf-actions}\\
\CAM & = \Lambda_1+\Lambda_2 +\Lambda_3.\nonumber
\end{align}
We call \scalefactor\ the scaling parameter by analogy with the two-planet case \citep{Michtchenko2008}.
Here, \CAM\ is the circular and coplanar angular momentum and verifies $\CAM= \AM+\AMD$.
Using the method described in section \ref{sec:technicalsetup}, we can do a formal second-order averaging over $\theta_\scalefactor$ and $\theta_\CAM$ because these angles are not resonant.
The Hamiltonian takes the form 
\begin{equation}
\H = \H_0(\boldsymbol{\resact}) +\epsilon\bar\H_1(\boldsymbol{\resact}) +\epsilon^2 \bar\H_{2,\mathrm{res}}(\boldsymbol{\resact},\theres) +\O(\epsilon^3)
\label{eq:ham3zero}
,\end{equation}
where $\epsilon^2\bar\H_{2,\mathrm{res}}(\boldsymbol{\resact},\theres)$ is the Hamiltonian of eq. \eqref{eq:H2full} averaged over $\theta_\scalefactor$ and $\theta_\AM$, and  $\boldsymbol\resact$ represents all the actions defined in \eqref{eq:transf-actions}.
In eq. \eqref{eq:ham3zero}, we can neglect $\epsilon\bar\H_1(\boldsymbol{\resact})$ as it is small with respect to $\H_0$ and only accounts for a correction of the mean motions of order $\epsilon$.
It should be noted that $\scalefactor$ and $\CAM$ are integrals of motion of \eqref{eq:ham3zero}, up to terms of order $\O(\epsilon^3)$. As a result, the Hamiltonian only has one degree of freedom and is integrable.
Another consequence of the conservation of $\CAM$ is that the zeroth-order three-planet MMRs preserve the system AMD.
In particular, if a system is only affected by these resonances, initially circular orbits will remain circular.
As such behaviour is observed in numerical simulations before the late instability, this result confirms the decisive role of zeroth-order three-planet MMRs in driving the instability.

We consider small variations of the actions around the resonance. Let us denote $\resact = \resact_0 +\smallvar{\resact}$ where $\resact_0$ corresponds to the value of $\resact$ such that the system is on the resonance curve \eqref{eq:03plMMRperrat}. Similarly, we have $\Lambda_k = \Lambda_{k,0} +\smallvar{\Lambda_k}$.
In turn, we have 
\begin{align}
\smallvar{\Lambda_1} &=  p\smallvar{\resact},\nnb
\smallvar{\Lambda_2} &=  -(p+q)\smallvar{\resact},\label{eq:relsmallvarLambda}\\
\smallvar{\Lambda_3} &=  q\smallvar{\resact},\nonumber
\end{align}
and so the inner and outer planets are moving in the same direction while the middle planet is moving in the opposite direction.
At first order, we can express the change in the period ratio  \perratvar{12} and  \perratvar{23} as a function of $\smallvar{\resact}$,
\begin{align}
\perratvar{12} & = 3\perrat{12,0}\left(\frac{p}{\Lambda_{1,0}}+\frac{p+q}{\Lambda_{2,0}}\right)\smallvar{\resact},\nnb
\perratvar{23} & = -3\perrat{23,0}\left(\frac{p+q}{\Lambda_{2,0}}+\frac{q}{\Lambda_{3,0}}\right)\smallvar{\resact}.
\label{eq:smallvarperrat}
\end{align}
We can take the ratio of the small variations \perratvar{23} and \perratvar{12} , and using eq. \eqref{eq:03plMMRperrat} to replace $p$ and $q$ we obtain the differential equation
\begin{equation}
\deriv{\perrat{23}}{\perrat{12}} = -\frac{\perrat{23}}{\perrat{12}} \frac{m_2(1-\perrat{12}\perrat{23})+m_1\perrat{12}^{-1/3}(1-\perrat{12})}{m_2(1-\perrat{12}\perrat{23})+m_3\perrat{12}\perrat{23}^{1/3}(1-\perrat{12})},
\label{eq:diff-dir}
\end{equation}
which gives the direction of the change of period ratios anywhere in the plane (\perrat{12},\perrat{23}) due to the neighbouring resonance.
We note that the equation no longer depends explicitly on the integers $p$ and $q$. Indeed, while the strength of each resonance depends on the resonance index $2(p+q)$ (see the following section), the resonant motion direction can be extended as  a continuous function of the period ratios using the resonance locus equation \eqref{eq:03plMMRperrat}.

The solution of eq. \eqref{eq:diff-dir} gives the direction of the Chirikov diffusion if the system dynamics were entirely governed by the zeroth-order three-planet MMRs. The differential equation can be integrated numerically given an initial condition, and the solution for the system studied in section \ref{sec:pheno} is displayed in orange in Fig. \ref{fig:pheno-network}. We see that the system follows the
diffusion direction for the majority of its  lifetime, but leaves it as soon as the dynamics are no longer driven by the three-planet MMR network.
We note that the problem has been reduced to study the diffusion along a one-dimensional curve. 

\subsection{Explicit size of the resonance width}
\label{sec:network-resdyn}

From the previous section, we know that the dynamics of a single zeroth-order three-planet resonance are integrable. 
Provided that these resonances overlap, we also have seen along which curve the motion should take place.
However, there is no guarantee that the neighbouring three-planet MMRs interact.
If the resonances are well isolated because their width is smaller than their separation, there is no possibility for large-scale chaos. 
In this case, a system could be influenced by a single resonance and never jump to the other ones.
The system will be almost secular and in principle could be considered as long-term stable.
Moreover, the diffusion timescale is linked to the resonance strength. It is also possible that while the resonances are overlapped, the diffusion along the network is so slow that it is meaningless for astrophysical applications.

One therefore needs to study the dynamics in detail to evaluate the strength of the three-planet resonances. We therefore carry out in this section a detailed and quantitative derivation of the perturbative steps described above, keeping track of all the relevant terms which contribute to three-planet MMRs.
We limit ourselves to a leading-order computation. As a result, we only keep the terms that do not depend on the eccentricity in the final expression. We also neglect the indirect term of the perturbing Hamiltonian $\epsilon\H_1$ as its value only affects the resonances when either $p$ or $q$ is equal to 1.
It is instead necessary to keep terms to first order in eccentricity because they contribute to terms independently of the eccentricity at second order in mass. The terms of the perturbing Hamiltonian $\epsilon\H_1$ that we consider are therefore \citep{Laskar1995,Murray1999}
\begin{align}
\epsilon\H_1 =& \sum_{1\leq i<j\leq 3}\sum_{l\in\Z} W_{ij}^l \EXP^{\i l(\lambda_i-\lambda_j)}\label{eq:exprH1}\\
&+\sum_{1\leq i<j\leq 3}\sum_{l\geq0} \left(V_{ij,<}^{l}x_i+V_{ij,>}^{l}x_j\right) \EXP^{\i (l\lambda_i-(l+1)\lambda_j)} + c.c.,\nonumber
\end{align}
where $c.c.$ designates the complex conjugate of the second sum, and
\begin{align}
W_{ij}^l &=  -\frac{m_in_j\Lambda_j}{2m_0} \lapc{1/2}{l}{\alpha_{ij}},\label{eq:Wijl}\\
V_{ij,<}^{l} &= \frac{m_in_j\Lambda_j}{2m_0} \sqrt{\frac{2}{\Lambda_i}} \left(l+1+\frac{\alpha_{ij}}{2}\dpart{}{\alpha}\right)\lapc{1/2}{l}{\alpha_{ij}},\label{eq:Vijl<}\\
V_{ij,>}^{l} &= -\frac{m_in_j\Lambda_j}{2m_0} \sqrt{\frac{2}{\Lambda_j}} \left(l+\frac{1}{2}+\frac{\alpha_{ij}}{2}\dpart{}{\alpha}\right)\lapc{1/2}{l}{\alpha_{ij}},\label{eq:Vijl>}
\end{align}
where $\alpha_{ij}=a_i/a_j$ and $ \lapc{s}{l}{\alpha}$ are the Laplace coefficients.
We refer to Appendix \ref{app:Laplace coefficients} for a definition and study of the Laplace coefficients and how to approximate them.
Here we note that \cite{Quillen2011} used a simplified approximation that confers the advantage of being easy to manipulate
\begin{equation}
\lapc{1/2}{l}{\alpha} \simeq  \frac{1}{2}\left|{\ln(1-\alpha)}\right|\EXP^{-|l|(1-\alpha)}.
\label{eq:lapcoeffapprox-quillen}
\end{equation}
In the limit of $\alpha$ very close to 1, the asymptotic equivalent of the Laplace coefficients does not depend on $l$ \citep{Laskar1995}. However, we find that for a fixed $\alpha$, for large $|l|$, the Laplace coefficient is asymptotic to
\begin{equation}
	\lapc{1/2}{l}{\alpha} \simeq \frac{2\alpha^{|l|}}{\sqrt{\pi|l|(1-\alpha^2)}}.
	\label{eq:lapcoeffapprox-exactmain}
\end{equation}
In the close planet approximation, $\alpha\to 1^-$, we get
\begin{equation}
	\lapc{1/2}{l}{\alpha} \simeq  \sqrt{\frac{2}{\pi|l|(1-\alpha)}}\EXP^{-|l|(1-\alpha)}
	\label{eq:lapcoeffapprox-main}
.\end{equation}
 We refer to Appendix \ref{app:Laplace coefficients} for a detailed discussion. 

The exponential dependency on $-|l|(1-\alpha)$ of the Laplace coefficients has two consequences for $W_{ij}^l$.
The interactions between planets 1 and 3 are always negligible with respect to the interaction in adjacent pairs, that is,\ 
for a given $l$, $|W_{13}^l|\ll |W_{j,j+1}^l|$. Similarly, for a given resonance, higher order terms in the resonant angle such as $\EXP^{\i N\theres}$ for $N>1$ are always negligible. 
It should also be noted that the formal development \eqref{eq:exprH1} is possible because the infinite sum can be replaced by a truncated sum in the actual computation due to the exponential decay of the coefficients. As a result, the neglected terms can be moved into the remainder in $\O(\epsilon^3)$.

With these clarifications, let us return to the calculation of the perturbative steps, starting from the perturbative Hamiltonian \eqref{eq:exprH1}.
The solution $\epsilon \chi_1$ to the corresponding homological equation has for expression\footnote{Cfr.\ eq.\ \eqref{eq:chi}; note in the equation that the sum over $l$ does not contain the secular term $l=0$}
\begin{align}
\epsilon \chi_1 =& \sum_{1\leq i<j\leq 3}\sum_{l\in\Z^*} \frac{W_{ij}^l}{\i l(n_i-n_j)} \EXP^{\i l(\lambda_i-\lambda_j)}\label{eq:exprchi1}\\
&+\sum_{1\leq i<j\leq 3}\sum_{l\geq0} \frac{\left(V_{ij,<}^{l}x_i+V_{ij,>}^{l}x_j\right)}{\i (ln_i-(l+1)n_j)} \EXP^{\i (l\lambda_i-(l+1)\lambda_j)} + c.c.,\nonumber  
\end{align}
where $\Z^*=\Z\setminus\{0\}$.
One can note that because of the small denominators in both sums, the coordinate transformation is not valid (\ie\  not close to the identity) in the vicinity of the co-orbital resonance (first sum) and the first-order two-planet MMRs (second sum).
As explained schematically in the previous section, this perturbation step produces $\mathcal O(\epsilon^2)$ terms, which we now calculate explicitly. In essence, we must only calculate the term $\epsilon^2\bar\H_2$ in \eqref{eq:ham3zero}. Since we would subsequently average over $\theta_\CAM$ and $\theta_\scalefactor$, the only terms that remain in $\epsilon^2 \bar\H_2$ must depend on the angle $(p\lambda_1 -(p+q)\lambda_2 +q\lambda_3)$ or its opposite. 
Because of the form of $\epsilon \chi_1$ and $\epsilon \H_1$, the only terms contributing at zeroth order in eccentricity to the averaged Hamiltonian $\epsilon^2\bar\H_2$ are of the form 
\begin{align}
&\left\{ \frac{W_{12}^p}{\i p(n_1-n_2)} \EXP^{\i p(\lambda_1-\lambda_2)},W_{23}^{q} \EXP^{-\i q(\lambda_{2}-\lambda_{3})}\right\},\nnb
&\left\{ \frac{W_{23}^{q}}{-\i q(n_2-n_3)} \EXP^{-\i q(\lambda_{2}-\lambda_{3})},W_{12}^p \EXP^{\i p(\lambda_1-\lambda_2)}\right\},\\
&\left\{\frac{V_{12,>}^{p}}{\i (pn_1-(p+1)n_2)}x_2 \EXP^{\i (p\lambda_1-(p+1)\lambda_2)},V_{23,<}^{q-1}\bar{x}_2\EXP^{-\i ((q-1)\lambda_2-q\lambda_3)}\right\}\nnb
&\left\{\frac{V_{23,<}^{q-1}}{-\i ((q-1)n_2-qn_3)}\bar{x}_2 \EXP^{-\i ((q-1)\lambda_2-q\lambda_3)},V_{12,>}^{p}x_2\EXP^{\i (p\lambda_1-(p+1)\lambda_2)}\right\}\nonumber
\end{align}
or their complex conjugates.
We note that in all the considered terms, the Poisson bracket can be reduced to the derivations with respect to the coordinates $\Lambda_2$, $\lambda_2,$ and $x_2$ of planet 2 only, as they are the only ones to appear on both sides.
For the two last terms, we only keep the terms where the eccentricity dependency is dropped due to the derivation operator $\i \dpart{\cdot}{x_2}\dpart{\cdot}{\bar{x}_2}-\i\dpart{\cdot}{\bar{x}_2}\dpart{\cdot}{x_2}$.
This gives
\begin{equation}
\epsilon^2 \bar\H_2 = \epsilon^2 \rescoefpq\cos(\theres),
\label{eq:H2first-order}
\end{equation}
where
\begin{align}
\epsilon^2 \rescoefpq =& -\left(\frac{q}{p(n_1-n_2)}+\frac{1}{n_2-n_3}\right)\dpart{W_{12}^{p}}{\Lambda_2}W_{23}^q\nnb
& +\left(\frac{1}{n_1-n_2}+\frac{p}{q(n_2-n_3)}\right)W_{12}^{p}\dpart{W_{23}^q}{\Lambda_2}\nnb
& - \left(\frac{q}{p(n_1-n_2)^2}+\frac{p}{q(n_2-n_3)^2}\right)W_{12}^{p}W_{23}^q\dpart{n_2}{\Lambda_2}\\
\nonumber
&+\left(\frac{1}{pn_1-(p+1)n_2}+\frac{1}{(q-1)n_2-qn_3}\right)V_{12,>}^{p}V_{23,<}^{q-1}.
\end{align}
Because $\epsilon^2 \bar\H_2$ is small with respect to the Keplerian part, we evaluate the action at the nominal resonance value, that is, $pn_1-(p+q)n_2+qn_3=0$. Using eqs. \eqref{eq:03pMMR}, \eqref{eq:perrat}, \eqref{eq:Wijl}, \eqref{eq:Vijl<}, \eqref{eq:Vijl>} as well as
\begin{equation}
\dpart{\alpha_{12}}{\Lambda_2} = -2\frac{\alpha_{12}}{\Lambda_2}, \quad \dpart{\alpha_{23}}{\Lambda_2} = 2\frac{\alpha_{23}}{\Lambda_2} \text{ and } \dpart{n_2}{\Lambda_2} = -3\frac{n_2}{\Lambda_2},
\end{equation}
we can express $\epsilon^2 \rescoefpq$ as 
\begin{align}
\epsilon^2 \rescoefpq =& \frac{m_1m_3n_2\Lambda_2\alpha_{23}}{m_0^2\genperrat}
\left[(1-\resloc)b_{1/2}^{(q)}\left(1+\alpha_{12}\dpart{}{\alpha}\right)b_{1/2}^{(p)}\right.\nnb
&\phantom{\frac{m_1m_3n_2\Lambda_2\alpha_{23}}{m_0^2}	\left[\right.} +\resloc\alpha_{23}b_{1/2}^{(p)}\dpart{b_{1/2}^{(q)}}{\alpha}\label{eq:Rpqexact}\\
&\phantom{\frac{m_1m_3n_2\Lambda_2\alpha_{23}}{m_0^2}	\left[\right.} \left.+\frac{3\resloc(1-\resloc)}{2\genperrat}b_{1/2}^{(p)}b_{1/2}^{(q)}\right]+\nnb
& \frac{m_1m_3n_2\Lambda_2\alpha_{23}}{m_0^2(\genperrat-(p+q)^{-1})}
\left[\frac{1-\resloc}{2}b_{1/2}^{(q)}\left(1+\alpha_{12}\dpart{}{\alpha}\right)b_{1/2}^{(p)}\right.\nnb
&\phantom{\frac{m_1m_3n_2\Lambda_2\alpha_{23}}{m_0^2}	\left[\right.} +\left(\frac{\resloc}{2}+\frac{1}{4(p+q)}\right)\alpha_{23}b_{1/2}^{(p)}\dpart{b_{1/2}^{(q)}}{\alpha}\nnb
&\phantom{\frac{m_1m_3n_2\Lambda_2\alpha_{23}}{m_0^2}	\left[\right.} +\frac{\alpha_{12}\alpha_{23}}{4(p+q)}\dpart{b_{1/2}^{(q)}}{\alpha}\dpart{b_{1/2}^{(p)}}{\alpha}\nnb
&\phantom{\frac{m_1m_3n_2\Lambda_2\alpha_{23}}{m_0^2}	\left[\right.} \left.+\resloc(1-\resloc)(p+q)b_{1/2}^{(p)}b_{1/2}^{(q)}\right],\nonumber
\end{align}
where the Laplace coefficients depending on $p$ (resp. $q$) are evaluated at $\alpha_{12}$ (resp. $\alpha_{23}$).

In this expression, the second prefactor can go to infinity for $\genperrat = 1/(p+q)$.
For the resonance defined by $p$ and $q$, this happens at the intersection of the two-planet MMRs $\perrat{12}=p/(p+1)$ and $\perrat{23}=(q-1)/q$.
This result is a consequence of the non-validity of the second-order averaging very close to a two-planet MMR.
Since we primarily focus on the regions outside of two-planet MMR, we ignore this feature in the following developments.
Moreover, for large $p+q$, the MMR intersections are within the region of the two-planet MMR overlap.

Under the assumptions made so far, the above expression is exact, and can be used for numerical explorations of the size and typical frequency of each three-planet resonance without impediment (see below).
To obtain further analytical insight, it is however quite cumbersome, and it does not clearly show which parameters of the planetary system and of the resonance play a role in determining the properties of the resonant motion.
We therefore aim to simplify the above expression, keeping always in mind that we are ultimately interested in the diffusion in period ratio space driven by these three-planet resonances, and specifically in the timescale that is needed for large-scale diffusion.
It is expected that the resonances with highest index dominate this timescale (see also below); in the remainder of this section we therefore take the limit $1/(p+q)\to0$ and expand around this value.
We note moreover that for $1/(p+q)\to0$, the second term in \eqref{eq:Rpqexact} blows up when $\genperrat\to0$, in which case also the first term would go to infinity; however $\genperrat\to0$ only happens when one of the period ratios $\perrat{i,i+1}\simeq1$: this limit is beyond the scope of the study, and so we can exclude this case.

With these considerations in mind, the above expression can be considerably simplified. To this end, we  make the close planet approximation: $1-\alpha_{ij}\ll 1$. 
We define 
\begin{equation}
\plsep_{ij} = 1-\alpha_{ij} \simeq \frac{2}{3}(1-\perrat{ij}),
\label{eq:plsep-ij}
\end{equation}
which is an excellent estimate for period ratios from 0.5 to 1.

The product of the Laplace coefficients and their derivatives in $\rescoefpq$ introduces an exponential factor of the form $\EXP^{-p\plsep_{12}-q\plsep_{23}}$ 
(cf. eq. \ref{eq:lapcoeffapprox-main})
that sets the order of magnitude of the resonance term.
We can therefore simplify expression \eqref{eq:Rpqexact} by taking advantage of the resonance relationship.
Indeed, for tightly packed systems, and in the vicinity of a resonance defined by $p$ and $q$, eq. \eqref{eq:03plMMRperrat} can be transformed into a relationship on the planet spacings:
\begin{equation}
p\plsep_{12} \simeq q \plsep_{23}.\label{eq:reseqspacing}
\end{equation}
By analogy with the generalised period ratio separation $\genperrat$,  we define a generalised orbital spacing that we note 
\begin{equation}
	\plsep = \frac{\plsep_{12}\plsep_{23}}{\plsep_{12}+\plsep_{23}} \simeq  \frac{p}{p+q}\plsep_{12} \simeq \frac{q}{p+q}\plsep_{23},
	\label{eq:plsep-def}
\end{equation}
where the two last equalities are approximations using eq. \eqref{eq:reseqspacing}.
We have 
\begin{equation}
\genperrat \simeq \frac{3}{2}\delta.
\end{equation}

Using the newly defined variable $\plsep$ and $\resloc$, the expression  of the coefficient $\epsilon^2\rescoefpq$ can be simplified to
\begin{equation}
\epsilon^2\rescoefpq = \frac{m_1m_3n_2\Lambda_2}{3\pi m_0^2}\frac{\resloc(1-\resloc)}{\plsep^2} \left(17+\frac{21}{(p+q)\plsep}\right)\EXP^{-2(p+q)\plsep},
\label{eq:Rpqapprox2}
\end{equation}
where we only keep the terms up to second order in $(p+q)\plsep$.
Since the exponential factor depends on $(p+q)\plsep$, the resonance mainly matters in the region where $(p+q)\plsep$ is of order unity, hence we choose to set $17+21/((p+q)\plsep))\simeq38$, which is approximately the value taken for $(p+q)\plsep\simeq1$ as it allows us to carry the computations analytically. This value also gives a more accurate estimate for the  width of the resonances (see below).

The  expression obtained in \eqref{eq:Rpqapprox2} for the three-planet resonance perturbation Hamiltonian is remarkable. The resonance strength only depends on its index and not explicitly on $p$ and $q$.
This means that all resonances with the same index can be compared very easily. In other words, the network of zeroth-order three-planet MMRs can be partitioned into subnetworks consisting of resonances with the same index.

To fully describe the resonant dynamics, we now go back to equation \eqref{eq:ham3zero} (we recall that we can safely drop the term $\epsilon\bar\H_1$); we now expand the Keplerian part around the resonance centre 
\citep{Chirikov1979,Petit2017}.
This is more easily done in the original Delaunay variables $\boldsymbol\Lambda$, and we have
\begin{equation}
\H_0 
\simeq \sum_{j=1}^{3}\left(-\frac{\mu^2m_j^3}{2\Lambda_{j,0}^2} + n_{j,0}(\Lambda_{j}-\Lambda_{0,j})- \frac{3n_{j,0}}{2\Lambda_{j,0}}(\Lambda_{j}-\Lambda_{0,j})^2 \right),
\end{equation} 
where the constant terms can be safely dropped.
Using Eqs. \eqref{eq:03pMMR} and \eqref{eq:relsmallvarLambda}, the first-order term vanishes and the coefficient of the second-order term has for expression
\begin{equation}
-\frac{\K_2}{2}= -\frac{3n_{2}}{2\Lambda_{2}}(p+q)^2\left(1+\frac{m_2}{m_1}\frac{\eta^2}{\alpha_{12}^2}+\frac{m_2}{m_3}\alpha_{23}^{2}(1-\eta)^2\right).\label{eq:K2}
\end{equation}
We note that $\K_2$ only depends on the index of the resonance and weakly on the planet masses.
Passing finally to the resonant canonical variables \eqref{eq:transf-angles} and \eqref{eq:transf-actions}, the resonant Hamiltonian has for expression
\begin{equation}
\H_\mathrm{res} = -\frac{\K_2}{2}(\resact-\resact_0)^2 +\epsilon^2R_{pq} \cos \theres.\label{eq:resHam}
\end{equation}
This is the standard pendulum Hamiltonian. The width of the resonance in the action space  is given by the expression \citep[\eg][]{Ferraz-Mello2007}
\begin{equation}
\reswidth{\resact} = 2\epsilon\sqrt{\frac{2\rescoefpq}{\K_2}}.
\label{eq:width}
\end{equation}
The small oscillation frequency is given by
\begin{equation}
\boxed{\freqpq = \epsilon\sqrt{\K_2R_{pq}} = n_2 \epsilon M \numfacfreq\frac{\sqrt{\resloc(1-\resloc))}}{\plsep}(p+q)\EXP^{-(p+q)\plsep},}
\label{eq:freqpq}
\end{equation}
where $\numfacfreq = \sqrt{\frac{38}{\pi}} = 3.47$ 
and 
\begin{equation}
\epsilon \Mfac = \frac{\sqrt{m_1m_3+m_2m_3\eta^2\alpha_{12}^{-2}+m_1m_2\alpha_{23}^{2}(1-\eta)^2}}{m_0}\label{eq:Mfac}
\end{equation}
is the relevant  mass ratio for the studied problem. For equal mass and equal tight spacing we have $\epsilon \Mfac = \sqrt{\frac{3}{2}}\frac{m_p}{m_0}$.
Contrary to the two-planet case, we note that the relevant mass ratio is not reduced to the sum of the planet masses over the mass of the star.
It is also interesting to note that the expression remains meaningful even in the case where one of the planets is reduced to a test particle.

The resonances have a clearer geometrical interpretation in the period ratio space than in the action space, particularly when one needs to compare them.
We therefore compute the width of the resonances perpendicularly to the network, that is, the width in terms of the variable $\resloc$.
Using Eqs. \eqref{eq:resloc} and \eqref{eq:smallvarperrat} and some algebraic manipulation, we have
\begin{equation}
\deriv{\eta}{\resact} = \frac{\K_2}{n_2(p+q)}\frac{\resloc(1-\resloc)}{\genperrat}.
\end{equation}
This means that the width in terms of $\resloc$ can be estimated as
\begin{equation}
\boxed{\reseta = \frac{4\sqrt{2}\resloc(1-\resloc)\freqpq}{3(p+q)\plsep n_2}  = \numfacres\epsilon\Mfac \frac{(\resloc(1-\resloc))^{3/2}}{\plsep^{2}}\EXP^{-(p+q)\plsep}.}\hspace{-0.3cm}
\label{eq:reswidtheta}
\end{equation}

We have thus shown that the width of the resonances in the period ratio plane depends exponentially on the MMR index and the prefactor is a continuous function of the period ratios.
In particular, it seems important to compare resonances with the same index because of their similar geometry.

\subsection{Resonance overlap}
\label{sec:network-overlap}

We wish to determine the sections of the period ratio space $(\perrat{12},\perrat{23})$ where resonances overlap.
Because of the expression of resonance width \reseta, we see that the width of the resonances close to a given point $(\perrat{12},\perrat{23})$ mainly depends on the index $p+q$. It is therefore natural to consider the density of the resonances for a fixed value of $p+q$.

We denote $\density_{k}(\plsep,\resloc)$, the local filling factor of the zeroth-order three-planet MMRs of index $\resind=p+q$. Here,
$\density_{\resind}$ corresponds to the proportion of the period ratio space occupied by this subnetwork.
Let us also define $\denstot(\plsep,\resloc)$, the filling factor of all zeroth-order three-planet MMRs. 
The filling factor measures the space locally\footnote{Here and later, by \emph{locally}, we mean a region large enough to contain resonances of different indexes such that the exact resonance position is not relevant, but small enough such that $\plsep$ and $\resloc$ do not vary significantly. A good example is a rectangle delimited by adjacent two-planet first-order MMRs.} occupied by all the nearby resonances of arbitrary index with respect to the available space in the period ratio plane. 
If \denstot\ is larger than 1, then there are enough resonances to locally cover the period ratio plane.

Such a filling factor is introduced by \cite{Quillen2011} for the same problem. However, these latter authors only consider resonances such that $|p-q|\simeq1$, which leads to  the exponential dependency being neglected.
We show here that taking into account all the resonances is critical to obtain an accurate diffusion rate and survival time.
The idea to count all the resonances without taking care of their precise position in order to obtain Chirikov's overlap estimate was also used with success for two-planet MMRs of arbitrary order \citep{Hadden2018}.

We have $\denstot \leq \sum_k \density_k$ since some resonances are counted multiple times. Indeed, if $p$ and $q$ are not coprime, the resonance lies on top of a resonance of lower index\footnote{More precisely, the index of the largest resonance is $(p+q)/\gcd(p,q)$.}.
Nevertheless, the contribution of a  resonance defined by two integers of the form $Np,Nq$  is negligible with respect to the contribution of the resonance $p,q$ because of the exponential decrease.
As a result, we consider that the overall resonance filling factor $\denstot$, is the sum of the subnetwork  filling factors $\density_{p+q}$.

Let us consider the subnetwork of resonances with index\footnote{
Technically, we refer to $2(p+q)$ as the index of the resonance; however the relevant quantities depend here on $p+q$ rather than $2(p+q)$.
We therefore refer to $p+q$ here as the index of the resonance for simplicity.} ${\resind = p+q}$.
The distance between two resonances in terms of \resloc\ is constant.
Indeed, let us consider the resonance defined by integers $p$ and $q$; its upper neighbour is defined by the integers $p+1$ and $q-1$, hence
\begin{equation}
	\resloc_{p+1,q-1} - \resloc_{p,q}= \frac{1}{p+q}=\frac{1}{k}.
\end{equation}
The filling factor $\density_{\resind}$ for the subnetwork of resonances with index $\resind$ can be determined by taking the ratio of the resonance width in terms of $\resloc$ with the distance between two neighbouring resonances in $\resloc$, that is,
\begin{equation}
	\density_{\resind} = k\reseta =\epsilon\Mfac \frac{(\resloc(1-\resloc))^{3/2}}{\plsep^{2}}\resind \EXP^{-\resind\plsep}.
\label{eq:density_k}
\end{equation}
The filling factor $\density_{\resind}$ thus depends on the subnetwork index $\resind$,  the generalised orbit spacing $\plsep$, the masses, and the resonant locator $\resloc$.

We approximate the total resonance filling factor \denstot\ as the sum of the subnetwork ones. We thus have
\begin{align}
	\denstot &= \numfacres\epsilon\Mfac \frac{(\resloc(1-\resloc))^{3/2}}{\plsep^{2}} \int_{0}^{+\infty}\resind e^{-\resind\plsep}\d \resind\nnb
	&=\numfacres\epsilon\Mfac\frac{(\resloc(1-\resloc))^{3/2}}{\plsep^{4}},
\label{eq:denstot}
\end{align}
where we have replaced the sum by an integral. The computations are also possible using the infinite sums but they result in a more complicated expression with a very limited gain in accuracy. This approximation is also done by \cite{Quillen2011}.

As in \cite{Quillen2011}, we find that the filling factor depends linearly on the mass ratio and scales as $\plsep^{-4}$.
However, our expression is valid for an arbitrary spacing and mass distribution, as long as the system is tightly packed.
We confirm that the natural spacing rescaling for the problem is not the Hill radius \eqref{eq:Hillrad}, which scales as $\epsilon^{1/3}$, but rather a dependency on $\epsilon^{1/4}$.
In particular, assuming that $\Mfac$ and $\resloc$ are constant, we can define a critical spacing value $\plsepov$ such that the zeroth-order three-planet MMR network fills the entire space. Taking $\denstot=1$ and solving for $\plsep$, one obtains
\begin{equation}
	\boxed{\plsepov = \epsilon^{1/4} M^{1/4}(\numfacres)^{1/4}(\resloc(1-\resloc))^{3/8}.}
	\label{eq:plsep-ov}
\end{equation}
Here, $\plsepov$ is a function of the masses and $\resloc$.
We can rewrite the filling factor \denstot\ as a function of $\plsepov$ as a power law over $\plsep$
\begin{equation}
\denstot = \left(\frac{\plsep}{\plsepov}\right)^{-4}.
\label{eq:denspowerlaw}
\end{equation}

In the case of equal mass and spacing systems, Eq. \eqref{eq:plsep-ov} becomes
\begin{equation}
	\plsep_{\mathrm{ov,eq}} = 1.16\left(\frac{m_p}{m_0}\right)^{1/4}.
	\label{eq:plsep-ov-eq}
\end{equation}

We note that $\plsep_{\mathrm{ov,eq}}$ corresponds to the generalised spacing defined in eq. \eqref{eq:plsep-def}. The actual orbit spacing is equal to $2\plsep_{\mathrm{ov,eq}}$ in this case.
The overlap criterion obtained by \cite{Quillen2011} is similar to ours since the exponent $1/4$ makes the numerical factors almost equal.

\begin{figure}
	\includegraphics[width=\linewidth]{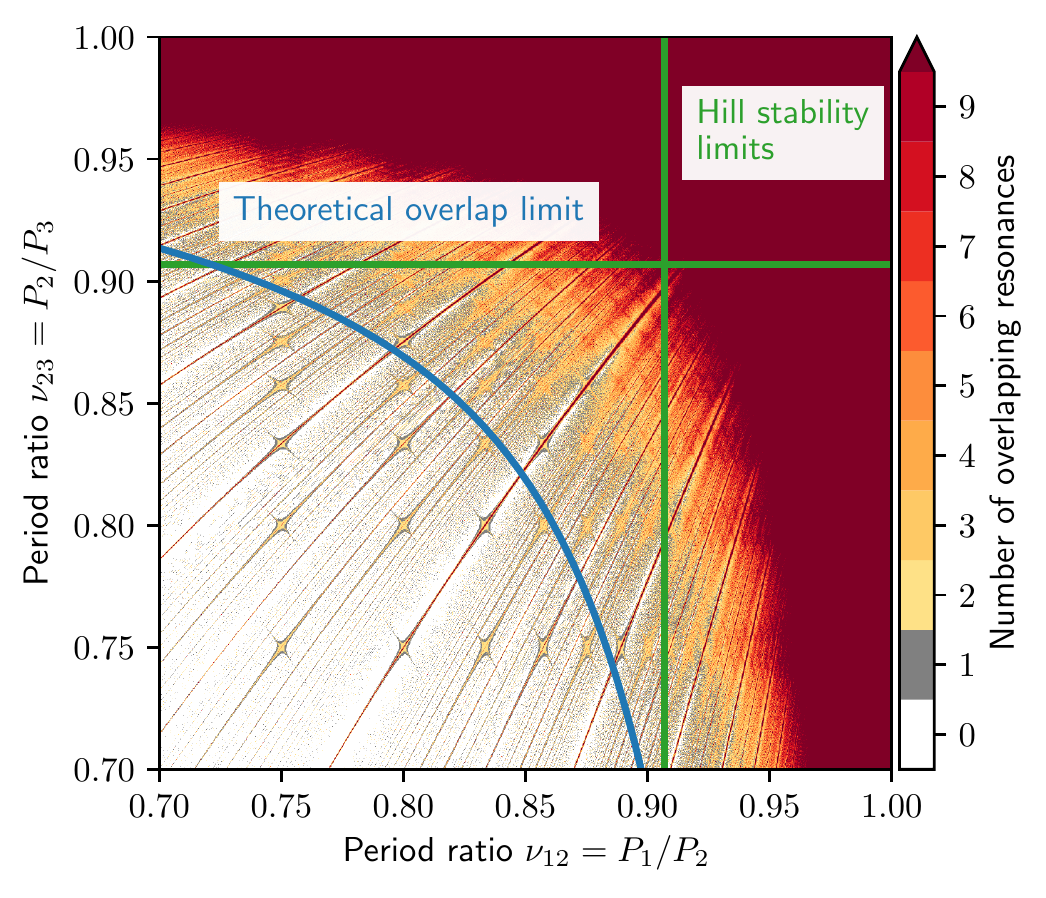}
        \caption{Number of resonances covering the period ratios space for equal-mass planets, with masses $10^{-5}$ \Ms.
	At first order, the number of resonances can be compared to the filling factor $\denstot$ (eq. \ref{eq:denstot}).
        We plot the two-planet circular Hill-stability limits \citep{Petit2018} for both planet pairs in green and the predicted overlap limit for the three-planet MMRs (eq. \ref{eq:plsep-ov}) in blue.
	\label{fig:overlap}}
\end{figure}

We plot in Figure \ref{fig:overlap}, the number of resonances that overlap at a given point in the plane $(\perrat{12},\perrat{23})$ for three equal-mass planets.
The mass of each planet is $10^{-5}$ \Ms.
The image is computed by creating a square grid of 2000 equally spaced period ratios between 0.7 and 1.
For each resonance index between 2 and 200, we compute for each point the closest resonance in terms of $\resloc$.
The closest resonance indicator is defined by $\resloc_{\mathrm{res}}$. 
The closest point on the resonance,  $(\perrat{12,\mathrm{res}},\perrat{23,\mathrm{res}})$, is found by gradient descent for the function $\resloc-\resloc_{\mathrm{res}}$.
We then compute the width in terms of $\resloc$ using eq. \eqref{eq:reswidtheta} at the point $(\perrat{12,\mathrm{res}},\perrat{23,\mathrm{res}})$ and compare it to the distance to the resonance $\resloc-\resloc_{\mathrm{res}}$.
We use the exact expression for $\rescoefpq$ \eqref{eq:Rpqexact}.
We count the resonances with multiplicities, that is,\ even if $p$ and $q$ are not coprime.
We see at the vicinity of the two-planet MMRs that the width of the three-planet resonances  increase due to the second term in eq. (\ref{eq:diffcoef}).  

The number of resonances is to first order a proxy for the filling factor \denstot\ (eq. \ref{eq:denstot}).
We see in Figure \ref{fig:overlap} that the region where the overlap of the three-planet MMR network takes place extends well beyond the Hill-stability limits (eq. \ref{eq:Hill}), particularly for comparable spacings between the two neighbouring planet pairs.
However, for very unequal spacings (away from the main lower-left to upper-right diagonal) we see that the overlap of only the three-planet MMRs is not enough to account for the instability and the two-planet interactions should be taken into account for the initial diffusion process.

\begin{figure}
	\includegraphics[width=\linewidth]{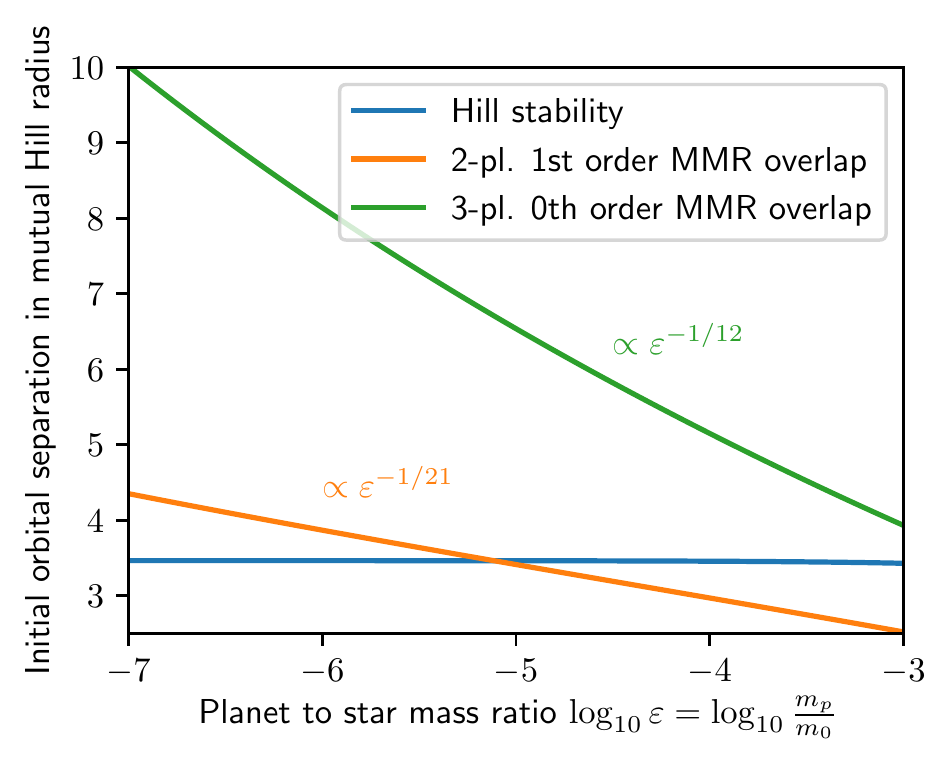}
        \caption{Comparison of the two-planet stability criteria, the Hill stability \citep{Gladman1993,Petit2018}, the first-order MMR overlap criterion, and the zeroth-order three-planet MMR overlap criterion expressed in units of the mutual Hill radius as a function of the planet-to-star-mass ratio for equal masses and equally spaced planets. \label{fig:compHillspacing}}
\end{figure}
To quantify how far the overlapping region extends, we consider systems of equally spaced planets with equal masses $m_p$ and plot in Figure \ref{fig:compHillspacing} the minimal spacing given by the Hill-stability limit \citep{Gladman1993,Petit2018}, the \cite{Wisdom1980} two-planet MMR overlap criterion, and our three-planet MMR overlap criterion (eq. \ref{eq:plsep-ov-eq}) as a function of the mutual Hill radius (eq. \ref{eq:Hillrad}).
We see that for small masses, the three-planet MMR overlap region goes to orbital spacing of the order 10 Hill radii, comparable to what was observed in previous numerical simulations.
It is also worth noting that we are only considering a restricted part of the resonance network.
Higher order three-planet resonances can also contribute to the filling factor even for circular orbits.
Moreover, for the larger masses, averaging to the second order in the masses as done in Sect. \ref{sec:technicalsetup} may not be sufficient.
Moreover, chaotic diffusion occurs in general well before the full overlap that is computed here \citep{Chirikov1979,Lichtenberg1992}.
Therefore, the phenomenon studied here could also work beyond the predicted limit.
 Our criterion should be seen as a lower limit.

Nevertheless, we predict that beyond a certain spacing, we should see an increase of the survival time approximately at the limit where the three-planet MMR network is not fully overlapped.
The limit should also appear at smaller separations in terms of Hill radius for larger planets (see section \ref{sec:EMScomp}) .

We have shown that there is a region where three-planet MMRs can contribute to a non-secular evolution, resulting in a diffusion in the semi-major axes of the planets, ultimately leading to the instability of the system when a first-order MMR is crossed.
The region can be determined quite accurately with the introduction of adapted variables and the computation of a resonance filling factor \denstot\ (eq. \ref{eq:denstot}).
We have seen that the resonance index is critical for determining the resonance width and that each equal-index subnetwork can be considered individually.
The minimum spacing between planets such that the resonance network does not cover the period ratio space scales as $\epsilon^{1/4}$ and not as the Hill radius.
However, we see that in some regions, the filling factor is well above 1, that is,\ only the smaller index MMRs are necessary to cover the space. 
As a consequence, the diffusion is faster when only wider resonances are involved, which will lead to the observed differences in \Tsurv. 

\section{Diffusion timescale}
\label{sec:timescale}
\subsection{Chirikov's diffusion}
\label{sec:diffusion-chirikov}

In the Chirikov model, a large-scale diffusion of the actions (or the frequencies) occurs when the resonances overlap.
Qualitatively, perturbations to the integrable resonance Hamiltonian \eqref{eq:resHam} create a stochastic layer in the vicinity of the separatrix.
When overlap occurs, the stochastic layers of adjacent resonances merge which allows diffusion along the network.
The diffusion rate depends on the resonance width and the period of the resonance.
One can estimate the diffusion rate for the resonance locator \resloc\ in the vicinity of a resonance as \citep{Chirikov1979}
\begin{equation}
	\diffcoef{p+q}=\frac{\freqpq}{2\pi}\left(\reseta\right)^2.
	\label{eq:diffcoef}
\end{equation}
This expression is an estimate that is valid\footnote{A more accurate expression can be derived by analysing the perturbations of Hamiltonian \eqref{eq:resHam}, we refer to \cite{Chirikov1979,Cincotta2002}.} when resonances overlap and the stochastic layers are well connected.
Studies of the Chirikov diffusion have been carried out for simplified Hamiltonian \citep{Giordano2004} or on astrophysical problems \citep{Cachucho2010}.
\cite{Cincotta2002} discusses the link between Chirikov and Arnold diffusion in astronomy and presents a modern description of Chirikov's theory.
The link with \cite{Nekhoroshev1977} theory is proposed in \citep{Cincotta2014}.
If the space is fully covered by overlapping resonances, the trajectory can be well approximated by a random walk.

The diffusion direction is perpendicular to the resonance in the action space.
This means that the diffusion is not isotropic, and to study the trajectory one would need to compute the contribution of every resonance to the diffusion tensor at every point. 
If the resonance lines do not intersect in the considered region, the diffusion will be well approximated by an unidimensional random walk perpendicular to the resonance network, with a negligible diffusion parallel to the resonances.
We can therefore consider a scalar diffusion coefficient given by the specific resonance that dominates the dynamics around the position in the phase space.
In particular, the diffusion coefficient is not constant and depends on the closest resonance width.
Such a diffusion corresponds to the behaviour observed in Fig. \ref{fig:pheno-network}.

In section \ref{sec:network-overlap}, we compute an overlap criterion taking into account every zeroth-order three-planet MMR.
However, as the resonance index increases, the associated diffusion rate vanishes, such that in the limit where the diffusion is dominated by smaller and smaller resonances, the timescale effectively tends to infinity.
However, for $\plsep<\plsepov$, not all the resonances are necessary to cover the phase space.
We therefore only need to consider the largest ones to compute the survival time.

\subsection{Partial resonance overlap}

We consider a small region around a point $(\plsep,\resloc)$ where the zeroth-order three-planet MMR network is locally overlapped.
Since $\denstot>1$, not all the resonances are necessary to cover the phase space. 
The largest resonances lead to the fastest diffusion, and therefore we need to only consider the subset of the widest resonances needed to cover the period ratio space in this given region. 
As the distance to reach a two-planet MMR is small, the main difference between the size of the resonances is governed by their index.
We therefore define an overlap index \ovind\ such that the subnetwork composed of the three-planet MMR with index $\resind$ smaller than \ovind\ locally covers the space.
Using Eqs. \eqref{eq:density_k} and \eqref{eq:denspowerlaw} we have
\begin{equation}
	\frac{\plsepov^4}{\plsep^2}\int_0^{\ovind} k \EXP^{-k\plsep}\d k = \left(\frac{\plsepov}{\plsep}\right)^4 \left[1-(\ovind\plsep+1)\EXP^{-\ovind\plsep}\right]= 1.
	\label{eq:defovind-implicit}
\end{equation}
We thus have an implicit definition of $\ovind$. We also note that the equation depends on $\ovind\plsep$ rather than \ovind\ alone. 
As a result, we define the variable
\begin{equation}
	\xiov = \ovind \plsep,
	\label{eq:xiovdef}
\end{equation}
which is a function of $\plsep/\plsepov$. There is no solution for Eq. \eqref{eq:defovind-implicit} in terms of elementary functions.
However, an explicit solution can be obtained using the Lambert $W$ function \citep[the function is also called inverse product log]{Corless1996}
\begin{equation}
	\xiov = -1-W_{-1}\left(\frac{\plsep^4-\plsepov^4}{\EXP\plsepov^4}\right),
	\label{eq:xiovexplicit}
\end{equation}
where $W_{-1}$ is the real valued branch with values smaller than $-\EXP$ defined between $-1$ and $0$. The function $W$ is the  inverse function to $z \to z\EXP^z$.
Using bounds on $W_{-1}$ by \cite{Chatzigeorgiou2013}, an excellent approximation of \xiov\ is
\begin{equation}
        \xiov= \sqrt{-2\ln\left(1-\frac{\plsep^4}{\plsepov^4}\right)}-\frac{2}{3}\ln\left(1-\frac{\plsep^4}{\plsepov^4}\right).%
\end{equation}

\subsection{Instability timescale}

Starting from a point $(\plsep,\resloc)$ in the period ratio space, we assume that the system wanders along the diffusion direction computed in section \ref{sec:network-MMRHamiltonian}.
While not exactly perpendicular to the resonant network in the period ratio space, the motion along the diffusion direction can be well parameterised by $\resloc$.
We therefore monitor the diffusion in terms of $\resloc$ because the resonance width and densities are easy to compute in terms of this variable.
Moreover, as seen in figure \ref{fig:pheno-network}, the systems are not too far from the first-order MMRs, and so the distances to cover are short and we can consider the period ratio as almost constant along the trajectory.

Since the diffusion coefficient \diffcoef{pq}\ mainly depends on $p+q$, one can associate a diffusion rate to each of the equal resonance index subnetworks.
We note $\etadist$, the distance in terms of $\resloc$ to the closest first-order MMR.
In order to describe the diffusion process, we adapt the framework developed by \cite{Morbidelli1997} to compute the escape rate of particles from the vicinity of invariant tori.
Let us assume that the system starts initially at a position $\resloc_0$, and becomes unstable once $\resloc$ reaches $\resloc_0+\etadist$.
Furthermore, we assume that the dynamics behave as a random walk.
The position of the system along the resonance network is described by the equation
\begin{equation}
	\deriv{\resloc}{t} = s(\resloc)b(t),
	\label{eq:mod-Langevin}
\end{equation}
where $s$ is related to the local diffusion coefficient and $b(t)$ is a Gaussian white noise with zero average, verifying
\begin{equation}
	\langle b(t)b(t')\rangle = 2\delta(t-t').
\end{equation}
For $s=s_0$ constant, Eq. \eqref{eq:mod-Langevin} is the classical Langevin equation. 
The associated diffusion equation for the probability density $p$ is
\begin{equation}
	\dpart{p}{t} = s_0^2\dpart{^2p}{\resloc^2},
\end{equation}
where the diffusion coefficient is $s_0^2$. By analogy with the case where $s$ is constant, we define $s(\resloc) =\sqrt{\diffcoef{p+q}}$, where $p$ and $q$ define the largest MMR that contains the point $\resloc$.
Since in the region considered, the resonances overlap, the interval $(\resloc_0,\resloc_0+\etadist)$ can be partitioned into the different subnetworks.
Each value of $\resloc$ can be associated with an index $\resind$ that corresponds to the widest resonance that contains it.
The probability that a given point $\resloc$ is in a resonance of index $k$ is given by
 \begin{equation}
	P\left(\resloc\text{ is in 3-pl.\ MMR of index }k\right) =  
	\begin{cases}
		\density_{k}\d k, & k\leq\ovind,\\
		0, &  k>\ovind
	\end{cases}.
	\label{eq:probres}
 \end{equation}
 We note that the considered value of $\resloc$ could also be contained in a higher index resonance than $\ovind$.
 However, the contribution of this higher order resonance to the diffusion is negligible.

 Following \cite{Morbidelli1997}, Eq. \eqref{eq:mod-Langevin} can be solved by introducing the variable
 \begin{equation}
	y(\resloc) = \int_{\resloc_0}^\resloc \frac{\d \resloc'}{s(\resloc')}.
	\label{eq:ydef}
 \end{equation}
Indeed, computing the time derivative of $y$ using Eqs. \eqref{eq:mod-Langevin} and \eqref{eq:ydef}, we have
\begin{equation}
	\deriv{y}{t} = \frac{1}{s(\resloc)}\deriv{\resloc}{t} = b(t),
\end{equation}
which is the Langevin equation with a unit diffusion coefficient.
The evolution of $y$ is therefore known and we obtain the evolution of $\resloc$ by inverting Eq. \eqref{eq:ydef}.
To do so, we need to determine the value of $s(\resloc)$.

Since we are interested in the overall diffusion speed and not the exact diffusion at a given point, we can attribute a probabilistic value to $s$ using eq. \eqref{eq:probres}.
We thus compute the average value of $y$ as a function of $\resloc$ over all the configurations of the resonance network. Noting $\bar y$ and $\bar \resloc$ the average values, we have
 \begin{align}
	\bar y(\resloc) &= \int_0^{\ovind} \int_{\resloc_0}^{\bar\resloc} y(\resloc')P(\resloc')\d \resloc' \d \resind 
	= \int_{\resloc_0}^{\bar\resloc}\d \resloc'\int_0^{\ovind} \frac{\density_k}{\sqrt{\diffcoef{k}}}\d k \nnb
	&= \left(\bar\resloc-\resloc_0\right)\int_0^{\ovind} \frac{\density_k}{\sqrt{\diffcoef{k}}}\d k. \label{eq:yexplicit}
 \end{align}
 We see that the variation of $\bar y$ is proportional to the variation of $\bar\resloc$, the integral being almost constant\footnote{In the sense that it does not depend on the exact structure of the resonance network.} around a given point of the period ratio space.
 In particular, we can define an effective diffusion coefficient, taking into account the contribution of all the resonances necessary to locally cover the phase space
 \begin{equation}
	\diffcoefeff = \left(\int_0^{\ovind} \frac{\density_k}{\sqrt{\diffcoef{k}}}\d k \right)^{-2}= 
	\left(\int_0^{\ovind} \frac{k}{\sqrt{\omega_k}}\d k \right)^{-2}.\label{eq:diffcoef1}
 \end{equation}
 Indeed, deriving Eq. \eqref{eq:yexplicit} gives 
 \begin{equation}
	 \deriv{\bar\resloc}{t} = \sqrt{\diffcoefeff}b(t).
 \end{equation}
We refer to Appendix \ref{app:Dawson} for the exact expression, which involves several special functions to evaluate the integrals.
It is nevertheless straightforward to evaluate the integrals numerically using \texttt{scipy} \citep{Virtanen2020} for instance.
We also provide a \texttt{jupyter} notebook reproducing the figures of this article\footnote{\url{https://github.com/acpetit/PlanetSysSurvivalTime}}.

For the remainder of the discussion, a very good estimate\footnote{The factor $10^{-\sqrt{-\ln(1-(\plsep/\plsepov)^4)}}$ was found by chance during exploratory work after having obtained the rest of the expression through power expansion for small $\plsep$. We have no clear explanation of its origin.} is obtained with the expression
\begin{equation}
		\diffcoefeff \simeq \epsilon\Mfac\numfacfreq n_2\frac{9\sqrt{\resloc(1-\resloc)}}{4\pi\sqrt{2}}\frac{\plsepov^{6}}{\plsep^4}\left(1-\frac{\plsep^4}{\plsepov^4}\right)10^{-\sqrt{-\ln(1-(\plsep/\plsepov)^4)}}.
		\label{eq:diffcoef-app}
\end{equation}
This expression is surprisingly compact, mainly depends on $\plsep$ and $\plsepov$, and does not contain an exponential term.
This expression is within an order of magnitude of the exact one over the majority of the considered range of $\plsep$.
We highlight the fact that the diffusion coefficient goes to zero for $\plsep\to\plsepov$.

We have reduced the initial problem to a simple continuous unidimensional random walk with constant diffusion coefficient.
The system can wander up until it reaches a first-order two-planet MMR on the diffusion direction.
The survival time distribution is thus given by the well-studied hitting time probability of a Brownian motion with two absorbing boundary conditions \citep{Redner2001,Borodin2002,Wax2009}.
We note $\resloc_{-}$ and $\resloc_{+}$ ($\resloc_{+}>\resloc_{-}$), the value of $\resloc$ at the neighbouring first-order two-planet MMRs along the diffusion direction.
The interval where the system can wander has for length $\Delta\resloc = \resloc_{+}-\resloc_{-}$.
The initial position on this interval can be measured by the quantity $u_0= (\resloc_0-\resloc_{-})/\Delta\resloc$ that is between 0 and 1.
\begin{figure}
	\includegraphics[width=\linewidth]{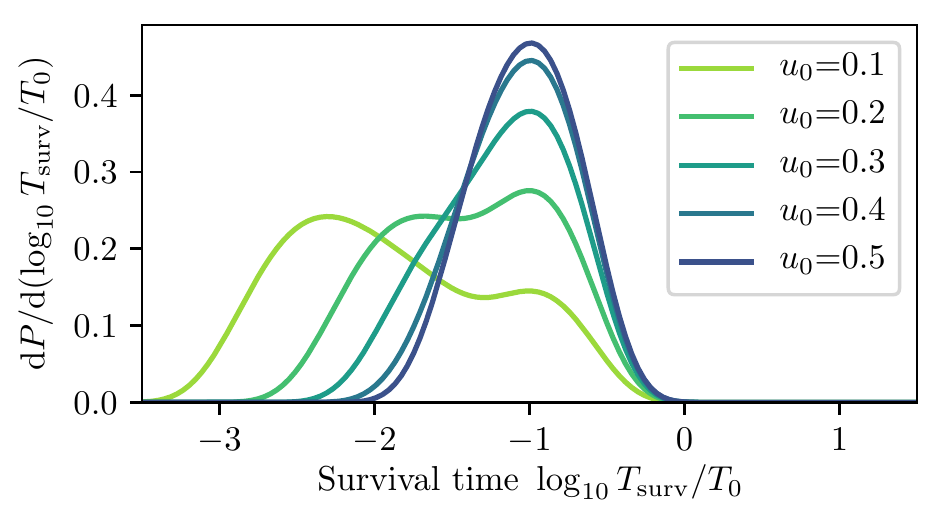}
	\includegraphics[width=\linewidth]{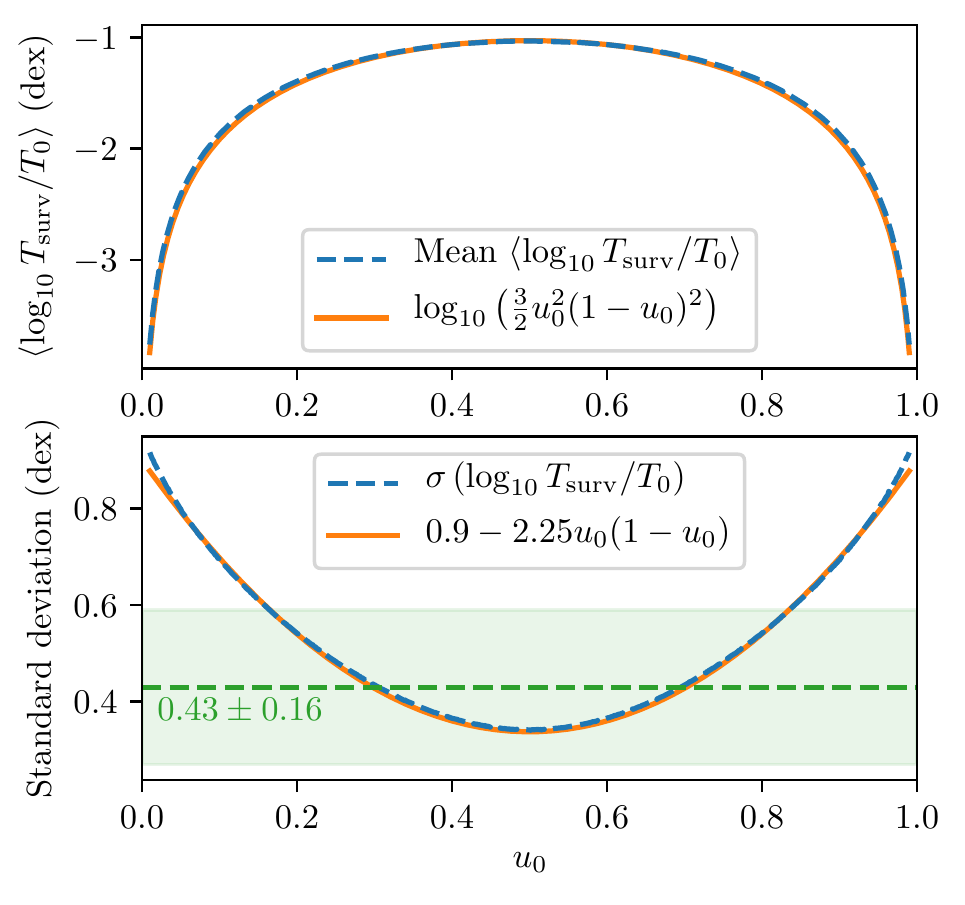}
        \caption{Top panel: Probability distribution function of $\log_{10}\Tsurv/T_0$ for various values of $u_0$, the normalised distance of the initial condition of the  system to the two neighbouring first-order two-planet MMRs. We see that for $u_0=0.5$, the time distribution is log-normal as observed in \cite{Hussain2020}. Bottom panels: Mean and standard deviation of the distribution as a function of $u_0$. We also plot approximate fits to the curves. We note that the mean remains within an order of magnitude of its maximum for almost all values of $u_0$. For $u_0$ close to 0.5, the standard deviation is close to the value $0.43\pm0.16$ measured by \cite{Hussain2020}, plotted as a green line with its 1$\sigma$ error. }\label{fig:Tsurvdist}
\end{figure}

The time can be rescaled such that the considered segment has length unity and the diffusion coefficient is 1 if we take as time unit
\begin{equation}
	T_0 = \frac{\Delta\resloc^2}{\diffcoefeff}.
	\label{eq:time-unit}
\end{equation}
The distribution of $\log_{10}\Tsurv/T_0$ is plotted in Fig. \ref{fig:Tsurvdist} as well as the mean and standard deviation of $\log_{10}\Tsurv/T_0$.
We give the expression of the distribution in Appendix \ref{app:specialfunction}.
As shown in Fig.~\ref{fig:Tsurvdist}, the mean of $\log_{10}\Tsurv/P_1$ can be well approximated as
\begin{equation}
	\left\langle \log_{10}\frac{\Tsurv}{P_1}\right\rangle = \log_{10}\frac{T_0}{P_1}+\log_{10}\left(\frac{3}{2}u_0^2(1-u_0)^2\right),
	\label{eq:meanlog}
\end{equation}
where we choose to normalise the survival time $\Tsurv$ by the innermost planet period $P_1$ to include $T_0$ in the right-hand side expression.
Similarly, we approximate the standard deviation as a second-order polynomial as
\begin{equation}
	\sigma\left( \log_{10}\frac{\Tsurv}{P_1}\right) = 0.9-2.25u_0(1-u_0).
	\label{eq:stdlog}
\end{equation}
We note that the standard deviation does not depend on the value of $\log_{10}T_0/P_1$ itself. Thus the standard deviation does not depend on the order of magnitude of the instability time.
This is a remarkable result, as it has been shown in numerical simulations \citep{Hussain2020} that the standard deviation of the survival time of extremely close initial conditions has the same properties.
Besides, \cite{Hussain2020} measured the standard deviation to be $0.43\pm0.16$ dex which is consistent with the value we obtain for initial conditions not too close initially to two-planets MMR, as can be seen in Figure \ref{fig:Tsurvdist}.

Equation \eqref{eq:meanlog} gives the expression of the mean survival time for any initial condition and while it involves terms that are not easily tractable analytically, they are easy to compute numerically given a specific system.
In the remaining part of this section, we seek to obtain a simplified expression to show how the mean survival time depends on the spacing and planet masses.

Let us assume that the system initially starts in between the two-planets MMRs $P$:$P$-1 and $P$+1:$P$ for the pair 1-2 and $Q$:$Q$-1 and $Q$+1:$Q$ for the pair 2-3.
In the example shown in section \ref{sec:pheno}, $P=Q=6$.
By evaluating $\resloc$ at the edges of the square created by the resonances, one can find that the maximum variation of $\resloc$ without encountering a MMR is
\begin{equation}
	\Delta\resloc_{\max} = \frac{P}{P+Q}-\frac{P-1}{P+Q} = \frac{1}{P+Q} = \genperrat,
\end{equation}
where the last equality is true for $\genperrat$ evaluated at the intersections of the MMRs  $P$:$P$-1 and $Q$+1:$Q$ or $P$+1:$P$ and $Q$:$Q$-1.
In practice, the variation of $\genperrat$ in a given resonance rectangle is $2\genperrat^2\ll\genperrat$ and so we neglect the variations of  $\genperrat$  in the rectangle.
Thus, using $\genperrat$ as the characteristic length for the diffusion interval of \resloc\ is accurate.
Here, $\resloc_{\max}$ describes the largest possible variation of $\resloc$, hence for any given point $(\perrat{12},\perrat{23})$.
For our simplified expression, we take $\Delta \resloc =\resloc_{\max}\simeq 3/2\plsep$.
Furthermore, we assume $u_0=0.5$. 
Using eq. \eqref{eq:diffcoef-app} to estimate the effective diffusion coefficient, we can get an order of magnitude for the survival time with the expression
\begin{empheq}[box=\fbox]{align}
	\left\langle \log_{10}\frac{\Tsurv}{P_1}\right\rangle \simeq& -\log_{10}\left(\frac{16\sqrt{2} A \epsilon\Mfac\sqrt{\resloc(1-\resloc)}}{3}\right)\nnb
	&+\log_{10}\left(\frac{\plsep^6}{\plsepov^6}\frac{1}{1-(\plsep/\plsepov)^4}\right)\label{eq:Tsurv_estimate}\\
	&+\sqrt{-\ln\left(1-\left(\frac{\plsep}{\plsepov}\right)^4\right)}.\nonumber
\end{empheq}
This expression can be decomposed into a prefactor that mainly depends on the planet-to-star-mass ratio and a function that only depends on $\plsep/\plsepov$.
At first glance, this expression is not linear in $\plsep$ which seems to contradict the numerical results \citep[\eg][]{Chambers1996,Obertas2017}, which hinted at a linear dependency of the logarithm of the survival time on the orbital spacing.
However, the function has an inflection point for $\plsep=0.629\plsepov$ and can be well approximated by a linear function of $\plsep/\plsepov$.
The linear approximation is correct in the regime of interest, that is,\ for Hill-stable planet pairs not too close to the overlap limit.
We have
\begin{equation}
\left\langle \log_{10}\frac{\Tsurv}{P_1}\right\rangle \simeq -\log_{10}\left( \epsilon \Mfac \sqrt{\resloc(1-\resloc)} \right) -  6.72  + 6.08 \frac{\plsep}{\plsepov}.
\label{eq:Tsurv_estimate_linear}
\end{equation}
To compare with previous numerical studies \citep{Chambers1996,Faber2007}, we compute the estimated survival time for equal-mass and equally spaced planets
\begin{equation}
	\left\langle \log_{10}\frac{\Tsurv}{P_1}\right\rangle \simeq -\log_{10}\left(\frac{m_p}{m_0}\right)-6.51+3.56 \left(\frac{m_0}{m_p}\right)^{1/4}\Delta.
\end{equation}
The slope coefficient 3.56 is very close to values obtained in previous works: \cite{Faber2007} estimated it at 3.7 and \cite{Yalinewich2020} at 3.4.
The prefactor proportional to $m_0/m_p$ is consistent with numerical simulations \citep{Chambers1996,Faber2007} and the numerical constant is very close to the one obtained by \citep{Faber2007}.

To summarise, we estimate the diffusion rate along the zeroth-order three-planet MMR network by only considering the widest resonances, up to the index $k=p+q$ such that they locally cover the period ratio plane.
We show that the complex random walk along the resonance network can be represented by a diffusion process with an effective locally constant diffusion coefficient given by Eq. \eqref{eq:diffcoef1}.
As observed in numerical simulations \citep{Hussain2020}, we show that the survival time distribution is approximately log-normal and we recover the same standard deviation.
Our estimation of the mean survival timescales as the planet separation in units of $\epsilon^{1/4}$ and not in units of Hill radii.
In particular, we emphasise the importance of considering  systems of various masses in such studies as it allows to discriminate between the physical mechanisms driving the dynamics.
Moreover, while our estimate is not exponential in planet spacing as fitted in numerical simulations, we show that for the range of times of interest, it can be considered as such.
As \cite{Quillen2011}, we predict that beyond the overlap limit, the survival time is likely much larger since the Chirikov diffusion is not an efficient process on its own.

\section{Comparison with numerical simulations}
\label{sec:EMScomp}

A large number of numerical studies have recently been performed on the problem of instability of tightly spaced planets \citep[\eg][]{Obertas2017}.
However, the most recent study limited itself to the minimal setup: an equal-mass and spacing (EMS) system with three planets on initially circular and coplanar orbits 
was performed by \cite{Faber2007}.
While these latter authors considered different mass ratios, the integration time was limited to $10^6$ inner planet orbits with a limited number of points.

\subsection{Equally spaced, equal-mass systems}

In order to have a fine enough resolution and a longer integration time, we run our own suite of numerical simulations.
We use \texttt{REBOUND} \citep{Rein2012a} and the symplectic integrators \texttt{WHFAST} \citep{Rein2015a} and \texttt{MERCURIUS} \citep{Rein2019a}.
We initialise three-planet systems on initially circular and coplanar orbits. 
As in previous studies \citep[\eg][]{Chambers1996}, the innermost planet semi-major axis is set to 1 au and the two outer-planet semi-major axes are chosen such that the two-planet pairs have an equal semi-major axis ratio. The initial angles are  randomly drawn.
We do three suites of simulations with planet masses of $10^{-7}\ \Ms$, $10^{-5}\ \Ms$, and $10^{-3}\ \Ms$. 
In each suite, the range of period ratios considered starts at 2 mutual Hill radii, that is,\ in the region where even the planet pairs should be considered unstable, and extends beyond the predicted three-planet MMR overlap criterion derived in section \ref{sec:network-overlap}.
Each system is integrated until two planets are closer than 1 Hill radius or until the inner planet has performed $10^9$ orbits.
For each of these system, we report the final time as the survival time. 
A value of 1 Gyr should therefore be understood as a lower limit.
We use \texttt{WHFAST} and a time-step of one-twentieth (1/20) of the inner orbit for the two smaller mass suites and \texttt{MERCURIUS} with a time-step of one-thirtieth (1/30) of the inner orbit for the largest mass planets.
We stopped running simulations at larger separations when it became clear that the systems were stable for 1 Gyr, outside of the first-order two-planet MMRs.
Each of the sets of simulations is composed of between 1600 and 1700 systems.

\begin{figure}
	\includegraphics[width=\linewidth]{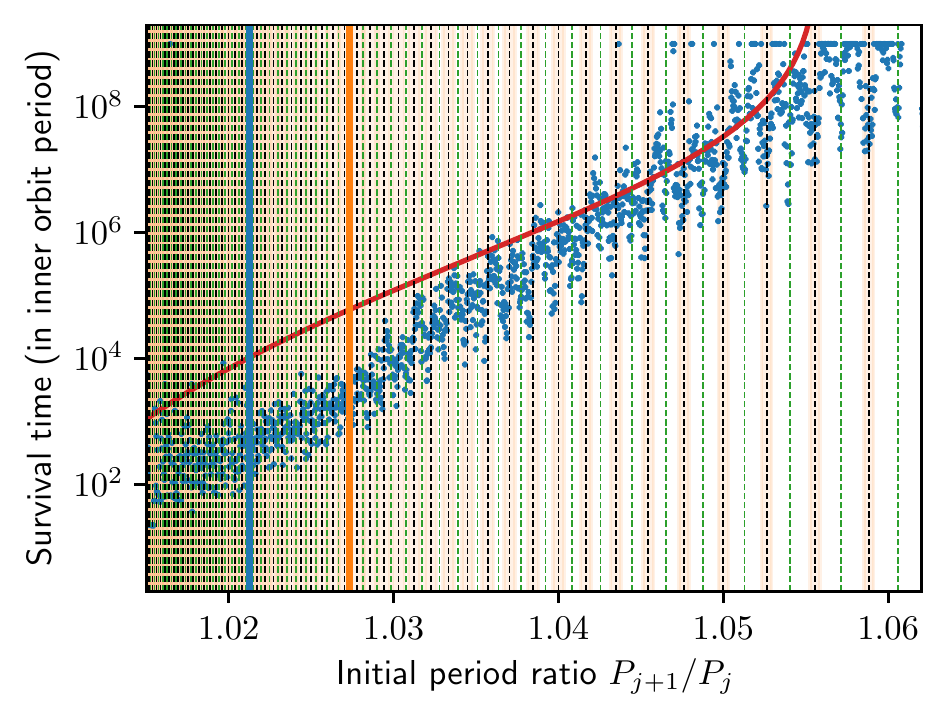}
        \caption{Survival time for a three-planet EMS system for planet masses of $10^{-7}\ \Ms$ as a function of the initial period ratio. \label{fig:EMS1e7} The red curve corresponds to the survival time estimate \eqref{eq:Tsurv_estimate}, the blue vertical line to the Hill-stability limit \citep{Petit2018}, the orange vertical line to the two-planet MMR overlap criterion \citep{Wisdom1980}. The dashed black (resp. green) lines are the two-planet first(resp. second)-order MMR. The light orange rectangles show an estimate of the width of the two-planet MMR \citep{Petit2017}.}
\end{figure}

\begin{figure}
	\includegraphics[width=\linewidth]{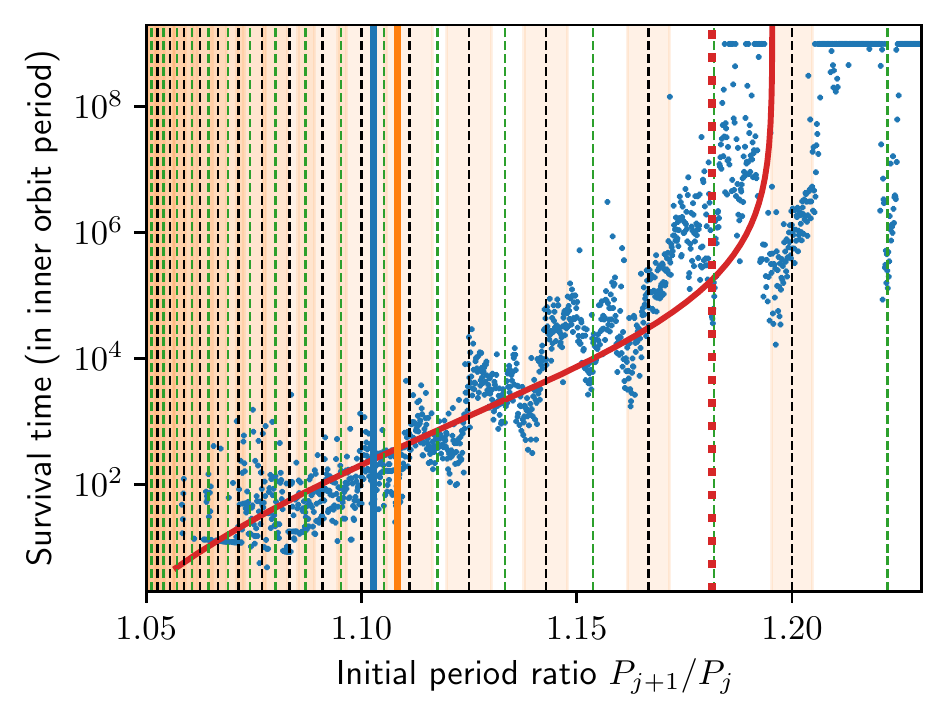}
	\caption{Survival time for a three EMS planet system for planet masses of $10^{-5}\ \Ms$ as a function of the initial period ratio. See Fig. \ref{fig:EMS1e7} for a detailed caption. The red dotted line corresponds to an alternate estimate position for the limit of the overlapped region (see text and eq. \ref{eq:altovcrit})\label{fig:EMS1e5}}
\end{figure}

\begin{figure}
	\includegraphics[width=\linewidth]{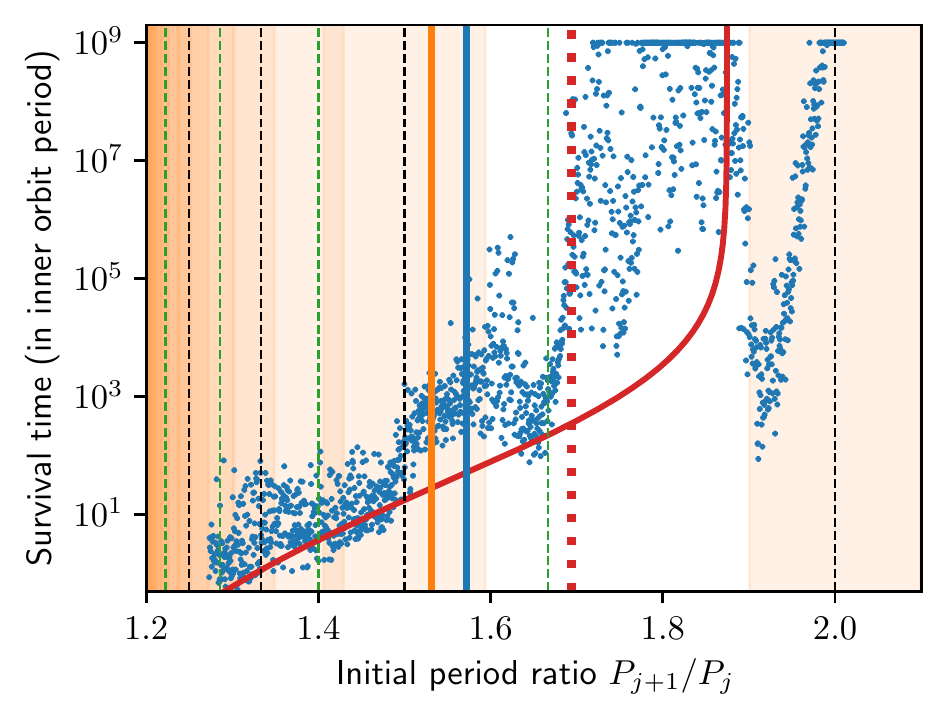}
	\caption{Survival time for a three EMS planet system for planet masses of $10^{-3}\ \Ms$ as a function of the initial period ratio. See Fig. \ref{fig:EMS1e7} for a detailed caption. The red dotted line corresponds to an alternate estimate position for the limit of the overlapped region (see text and eq. \ref{eq:altovcrit})\label{fig:EMS1e3}}
\end{figure}

We respectively plot in Figures \ref{fig:EMS1e7}, \ref{fig:EMS1e5}, and \ref{fig:EMS1e3}, the survival times for the sets of simulations with planet masses of $10^{-7}\ \Ms$, $10^{-5}\ \Ms$, and $10^{-3}\ \Ms$ and the associated analytical prediction \eqref{eq:Tsurv_estimate}, as a function of the initial period ratio.
Additionally, we add various other features that help us to understand the patterns that emerge in the survival time curves.
We plot the Hill stability limit \citep{Petit2018}, the two-planet first-order MMR overlap criterion \citep{Wisdom1980}, the nominal position of two-planet first-order MMRs $p+1:p$ and of two-planet second-order MMRs $p+2:p$.
We also plot an estimate of the width of first-order MMR for initially circular orbits \citep{Petit2017}.

The first thing to be noticed from these figures is that the survival time estimate \eqref{eq:Tsurv_estimate} of the logarithm of the survival time is consistent with the numerical simulations in the range where the former is almost linear.
The agreement is particularly striking for the intermediate case ($10^{-5}\ \Ms$).
We discuss explanations for the discrepancies for the low-mass and Jupiter-mass planets below.
Moreover, the slope being correct in all three figures is another indication that the scaling in $\epsilon^{1/4}$ is more appropriate than renormalising the spacing by the Hill radius.

We focus on Figure \ref{fig:EMS1e5} to describe more precisely the different features that should be pointed out.
First, if one ignores the variations due to the proximity to the resonances, our estimate lies in the middle of the distribution of survival time from period ratios of 1.05 to the end of the region where we consider that the three-planet MMRs overlap, around 1.195.
This means that our criterion slightly underestimates the diffusion time, as in Eq. \ref{eq:Tsurv_estimate} we assume a maximum value for the distance to the two-planet MMR network.
Nevertheless, Chirikov diffusion correctly predicts the slope as well as the right order of magnitude for the survival time.

We then notice, as previous studies have \citep{Smith2009,Obertas2017}, that the substructure on the curve is very well explained by the two-planet first-order MMR. 
By considering the regions outside of first-order MMR and the regions inside  separately, we can see that the survival time when the dynamics are dominated by the two-planet MMR is roughly two to three orders of magnitude lower than outside of their influence.
We can explain why the survival time is shorter for period ratios just below the Keplerian resonance, by noting that due to the shape of the first-order MMR the unstable fixed point where the first-order MMR separatrices originate is situated on the left of the resonance in the figures.
We note that if we had introduced a fluctuation in the initial period ratio, the pattern would be much less clear.
We can also note that the second-order MMRs likely play a role in accelerating the diffusion.
A similar effect is also clearly observed regarding the 2:1 MMR in Fig. \ref{fig:EMS1e3} as well as in Fig. \ref{fig:EMS1e7} where the larger apparent spread is due the the very dense two-planet MMR network.

Finally, we focus on the region close to where we predict that the three-planet MMR stops overlapping (period ratios of about 1.19) in Fig. \ref{fig:EMS1e5}.
We see that the linear trend followed in the range 1.05 to 1.16 no longer holds due to some systems surviving longer than expected, in particular beyond 1 Gyr.
Moreover, the spread of the survival times increases instead of staying constant as shown by \cite{Hussain2020}.
One can note that a similar feature is also visible in the results of \cite{Obertas2017}, although the increase in the spread is less visible therein, most likely because these latter authors consider five-planet systems instead of three.
In particular, outside of the 6:5 and the 11:9 resonance, it appears that systems live much longer than one would have expected from extrapolating the linear trend fitted in previous numerical studies.
The same behaviour is also observed in Fig. \ref{fig:EMS1e3}.
However, the region where short-lived and long-lived systems coexist is much larger because of the larger size of high-order two-planet MMRs that are not taken into account in this analysis.

These two observations, namely the longer survival times and the increased spreading around the overlap limit, are consequences of our analytical derivation.
Indeed, beyond the overlap region, the Chirikov diffusion alone can no longer drive the instability over homogeneous regions of the phase space.
This does not mean that systems beyond our approximate overlap limit will live indefinitely.
However, the instability in these systems is driven by an alternative mechanism to the Chirikov diffusion considered here.
In particular, we have neglected the diffusion parallel to the resonances, that is\ the Arnold diffusion \citep{Cincotta2002}.
However, these alternate pathways to instability are most likely much slower.

We now look more in detail at the apparent discrepancies between the estimated survival time and the two extreme cases.
In Figure \ref{fig:EMS1e7}, we see that while the analytical curve shows good agreement beyond period ratios of 1.035, the numerical simulations have a much shorter survival time at very close separations.
We postulate that two unaccounted-for effects play a significant role in this regime.
First, for very small bodies, the two-planet MMR overlap limit extends beyond the Hill-stability limit \citep{Deck2013,Petit2018}.
In particular, Chirikov diffusion is also possible along the two-planet MMR network.
Since the first-order two-planet MMRs are of comparable size, which depends on roughly $\epsilon^{2/3}$ for circular orbits, this network has a faster diffusion timescale.
Besides, in the derivation of the resonant coefficient expression in Sect. \ref{sec:network-resdyn}, we neglect the interactions between the inner and the outermost planets.
While these interactions are orders of magnitude smaller for larger masses and separations, they may be taken into account for very close planets.
However, the region where these long-distance interactions matter is of little interest and this effect is most likely smaller than the already mentioned two-planet MMR overlap.
We also notice in Fig. \ref{fig:EMS1e7} that we are not able to resolve the difference of behaviour in the non-overlapping region.
Indeed, the survival time depends on $\epsilon^{-1}$ and as a result the transition away from the Chirikov regime occurs for times of about $10^{10}$ orbits.

In the case of Jupiter-mass planets ($10^{-3}\ \Ms$), we see that the survival times 
starts to spread from $10^4$ orbits to more than $10^9$ around a period ratio of 1.7 whereas we estimate the overlap region to extend up to 1.85.
Moreover, our analytical estimate is too low in the region where the spread of survival time is consistent with a uniform diffusion process (up until period ratio of 1.7).
We should first point out that the analytical results in this mass regime are at most an extrapolation.
Indeed, the perturbation theory used in Sect. \ref{sec:technicalsetup} diverges for large perturbations.
Moreover, the close planet approximation is no longer valid for large spacing.
In particular, in our approximation of the Laplace coefficients (Eqs. \ref{eq:lapcoeffapprox-exactmain}, \ref{eq:lapcoeffapprox-main}), we replaced $\alpha^l$ with $\EXP^{-l(1-\alpha)}$ to carry the computations analytically.
Since $1-\alpha<-\ln(\alpha)$, we overestimate the resonance width and thus the extent of the overlap region.
In order to get a closer estimate for the particular case of equally spaced planets, one can replace $1-\alpha$ with $-\ln(\alpha)$ in the expression of $\plsep$ (Eq. \ref{eq:plsep-def}) and then in the expression of the resonance density (Eq. \ref{eq:denstot}) in order to compute the period ratio where overlap occurs.
After some algebraic manipulation, we estimate the actual limit to the overlap region to be situated at
\begin{equation}
	\alpha_{\mathrm{ov}}^{\mathrm{est.}} = \EXP^{-2\plsepov}.
	\label{eq:altovcrit}
\end{equation}
We note that we still rely on formulas obtained with the close planet approximation and as such, this result remains an attempt to understand a discrepancy between the analytical curve and the numerical simulations.
Figures \ref{fig:EMS1e5} and \ref{fig:EMS1e3} show the estimate \eqref{eq:altovcrit} as a red dotted line.
We see that they lie almost exactly where some systems start surviving beyond $10^9$ orbits.

\subsection{A non-equally spaced case}
\label{sec:nonEMS}

\begin{figure}
	\includegraphics[width=\linewidth]{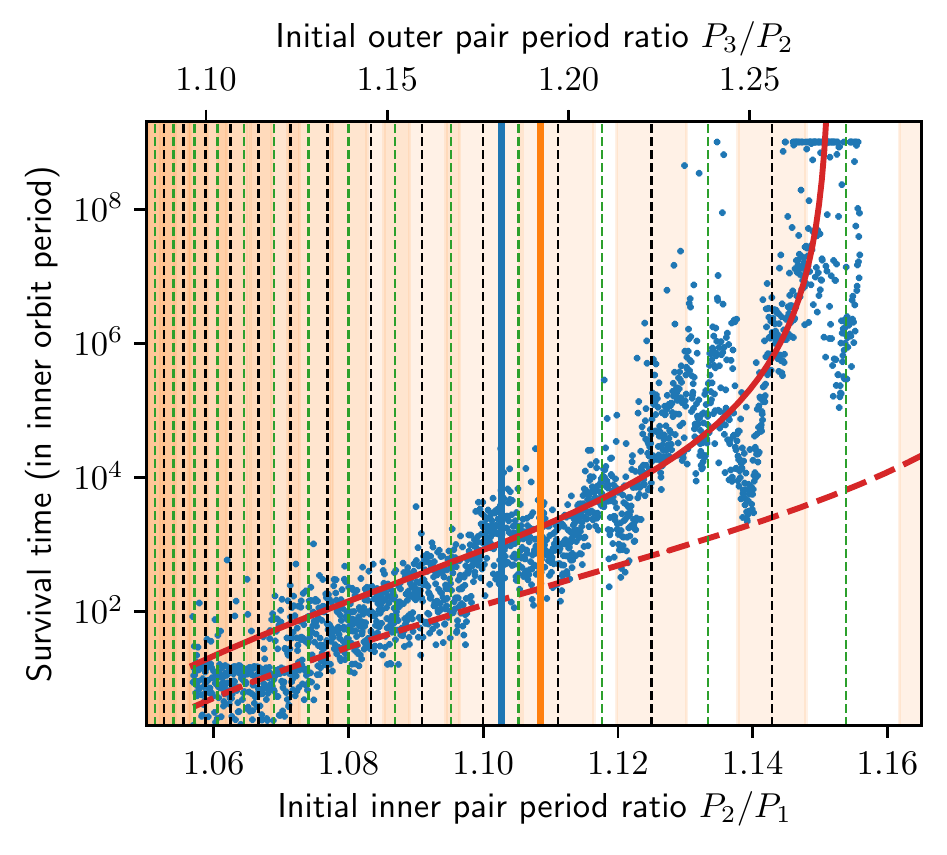}
        \caption{Survival time for non-equally spaced systems described in section \ref{sec:nonEMS} as a function of the initial inner-pair period ratio. The outer-pair period ratio is given at the top of the figure. The dashed survival-time estimate corresponds to the EMS case. See Fig. \ref{fig:EMS1e7} for a detailed caption.\label{fig:nonEMS1e5}}
\end{figure}

In order to demonstrate that our result is valid beyond the EMS case, we show an example of a configuration where the spacing between the inner and the outer pairs is different.
We initialise systems of three planets of equal mass ($10^{-5}\ \Ms)$ on initially circular and coplanar orbits.
For a given period ratio $\perrat{12}$ for the inner pair, we set the outer period ratio to
\begin{equation}
	\perrat{23} = 1-\phi(1-\perrat{12}),
\end{equation}
where $\phi=(1+\sqrt{5})/2\simeq1.68$ is the golden ratio and was chosen to avoid going through the intersections of two-planet MMRs, reducing their importance into the numerical results.
This leads to an outer pair spacing in period ratio that is $\phi$ times larger than for the inner pair.
The rest of the setup is similar to the EMS cases.

The survival time as a function of the initial inner period ratio is plotted in Figure \ref{fig:nonEMS1e5}.
The dashed red curve corresponds to the survival time for the EMS configuration shown in figure \ref{fig:EMS1e5}. 
We see that our model also successfully predicts the survival time in this case.
In particular, we show that the EMS estimate underestimates the survival time by roughly an order of magnitude.
The overlap limit is also reached at a tighter separation for the inner pair than in the equal spacing case.
Because there is no configuration where both planets are initialised close to a two-planet MMR, we observe less features in the the numerical results.

\subsection{Application to non-equal-mass planets}
\label{sec:TP}

\begin{figure}
	\includegraphics[width=\linewidth]{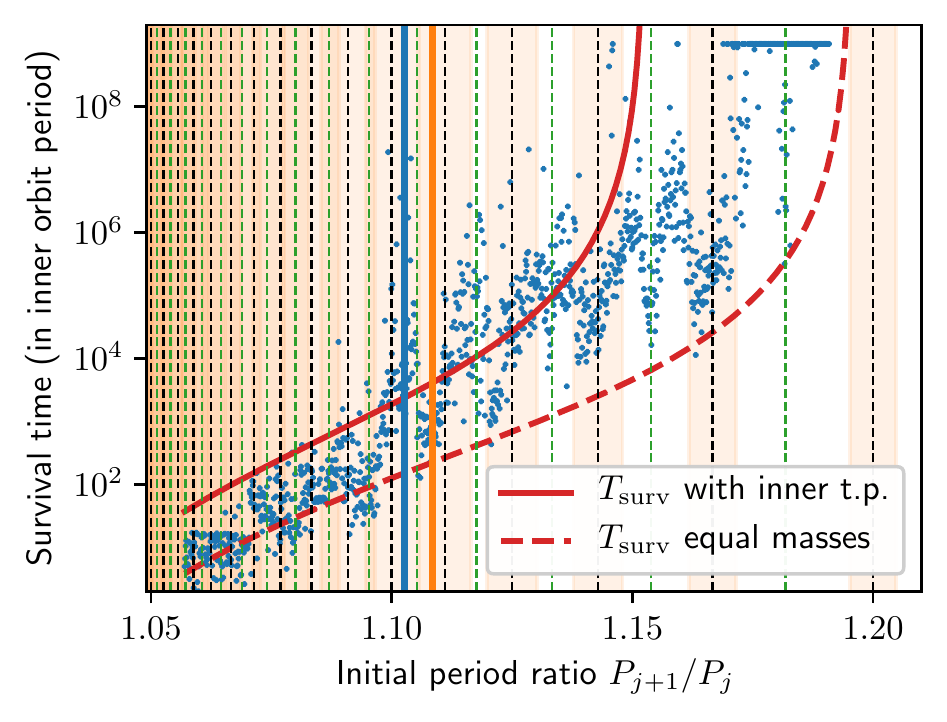}
        \caption{Survival time for a system composed of an inner test particle and two planets of equal mass $(10^{-5}\ \Ms$) with equal spacing as a function of the initial period ratio. See Fig. \ref{fig:EMS1e7} for a detailed caption. The full red curve is the prediction using the survival time estimate \eqref{eq:Tsurv_estimate} and the dashed line corresponds to the same estimate for the EMS case shown in figure \ref{fig:EMS1e5}. The red dotted line corresponds to an alternate estimate position for the limit of the overlapped region (see text and eq. \ref{eq:altovcrit})\label{fig:TP1}}
\end{figure}
\begin{figure}
	\includegraphics[width=\linewidth]{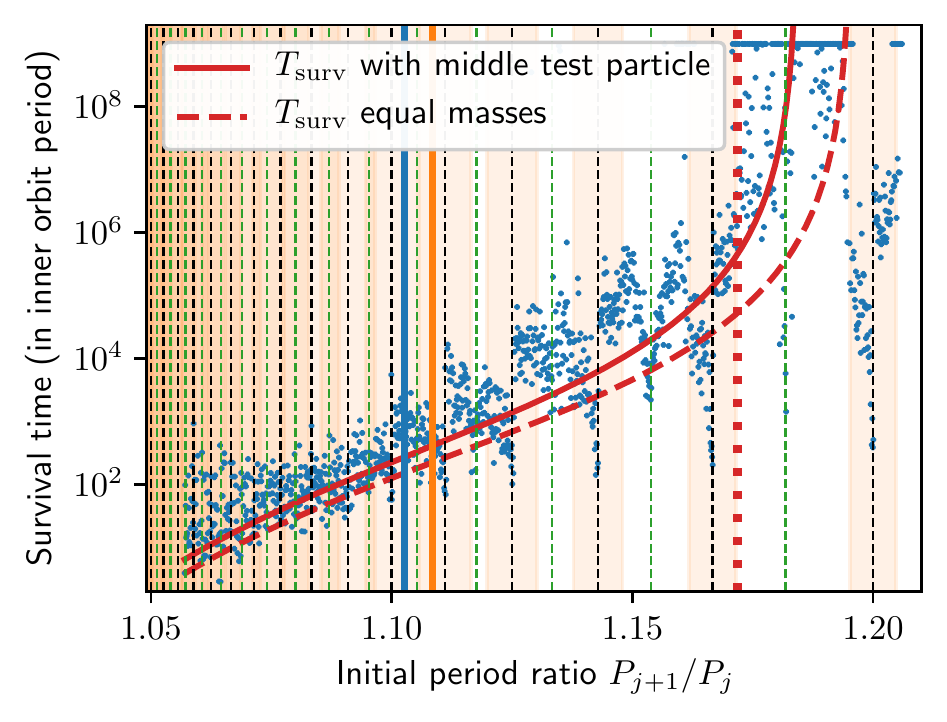}
	\caption{Same as figure \ref{fig:TP1} but with the test particle being placed between the two planets. \label{fig:TP2}}
\end{figure}

The relevant mass ratio introduced in our model, $\epsilon\Mfac$ (eq. \ref{eq:Mfac}), shows little variation in the case where one body has a significantly smaller mass than the two others.
Indeed, its value is maximised for three equal-mass planets for a given maximum planet mass $m_p$.
In this case, for equal spacing we get $\epsilon\Mfac\simeq1.22m_p/m_0$.
As already noted, $\epsilon\Mfac$ is non-zero as long as two planets are massive.
Assuming two equal-mass planets and a test particle, $\epsilon\Mfac=m_p/m_0$ if the test particle is in the middle and scales as $\resloc m_p/m_0$ (resp. $(1-\resloc) m_p/m_0$ ) if the test particle is the inner (resp. the outer) body.
In this case, $\epsilon\Mfac=m_p/m_0$ varies by less than a factor three between these extreme cases for initially equally spaced bodies.
To confirm this numerically, we plot on figures \ref{fig:TP1} and \ref{fig:TP2} the survival time of a test particle and two equal-mass planets in the case where the test particle is the inner and the middle body, respectively.
The massive planets have a mass of $10^{-5}\ \Ms$ and the remaining setup is similar to the one described in section \ref{sec:EMScomp}.
We plot with a dashed line the predicted time if the three planets were of equal mass.

The agreement is excellent in the case where the test particle is placed between the two planets (Fig. \ref{fig:TP2}). We recover the same features that were discussed in section \ref{sec:EMScomp}, including the corrected overlap limit (eq. \ref{eq:altovcrit}).
We also note that the predicted timescale is slightly longer than in the EMS case and this point seems confirmed by the numerical simulations.
When the inner planet is replaced by a test particle (Fig. \ref{fig:TP1}), we note that the estimate remains good before the predicted overlap but it seems that the overlap limit is underestimated in this case. 
We hypothesise that taking into account the interactions between the inner and the outer body may solve this discrepancy.

These numerical simulations confirm that diffusion along the three-planet MMR network is the main mechanism driving the instability for tightly packed systems.
In particular we highlight the very strong change of behaviour occurring at the limit of the fully overlapped region.
Moreover, the survival time estimate given by Chirikov diffusion accurately predicts the numerical simulation results over a wide range of survival times and the relevant range in orbital separations.

\section{Beyond three planets on circular orbits}
\label{sec:beyond}

\subsection{Systems with four or more planets}

As already seen in previous numerical studies, increasing the number of planets beyond three does not fundamentally change  the survival timescale.
\cite{Chambers1996} show that while there is a slight change in the slope of the survival time between systems of five planets and systems of three, the survival time is mostly unchanged by the addition of other planets into the systems\footnote{We note that \cite{Pichierri2020} show that adding more planets into a \emph{resonant} chain changes its stability. Further studies on this topic are required.}.

It is therefore natural to try  to extend our results to systems composed of more than three planets.
Unfortunately, contrary to the three-planet case, it is not possible to reduce the dynamics to a unidimensional Chirikov diffusion.
Indeed, the resonance network cannot be projected into a two-dimensional plane as done in section \ref{sec:network}.
One solution is to consider triplets of adjacent planets, and assume that this triplet is perturbed by the additional planets.
The influence of the other planets can be seen as a change in the period ratio $\perrat{12}$ and $\perrat{23}$ due to the resonances with the adjacent planets.
Assuming that the planet spacings are comparable, the perturbation of the period ratio is of the same order of magnitude as the one induced by the three-planet MMR from the considered triplet.

As a result, we can modify the resonance density $\density_k$ (eq. \ref{eq:density_k}) by including a multiplicative factor $K$ representing the influence of the other planets.
This is similar to assuming that the network is composed of $K$ times more resonances than previously accounted for.
The planets are mainly influenced by their direct neighbours.
Taking a conservative approach, we consider that both the inner and outer neighbours of the triplet increase the number of resonances affecting the three-planet subsystem  by 50\%. 
We therefore take $K=2$ as a reasonable guess.
The  overlap spacing of a system of five or more planets is given by $K^{1/4}\plsepov$.
The survival time is also affected because, while $\density_k$ has changed, the resonance width has not.
As a result, the effective diffusion coefficient (eq. \ref{eq:diffcoef1}) becomes $K^{-2}\diffcoefeff$, where the change of $\plsepov$ should be accounted for.

\begin{figure}
	\includegraphics[width=\linewidth]{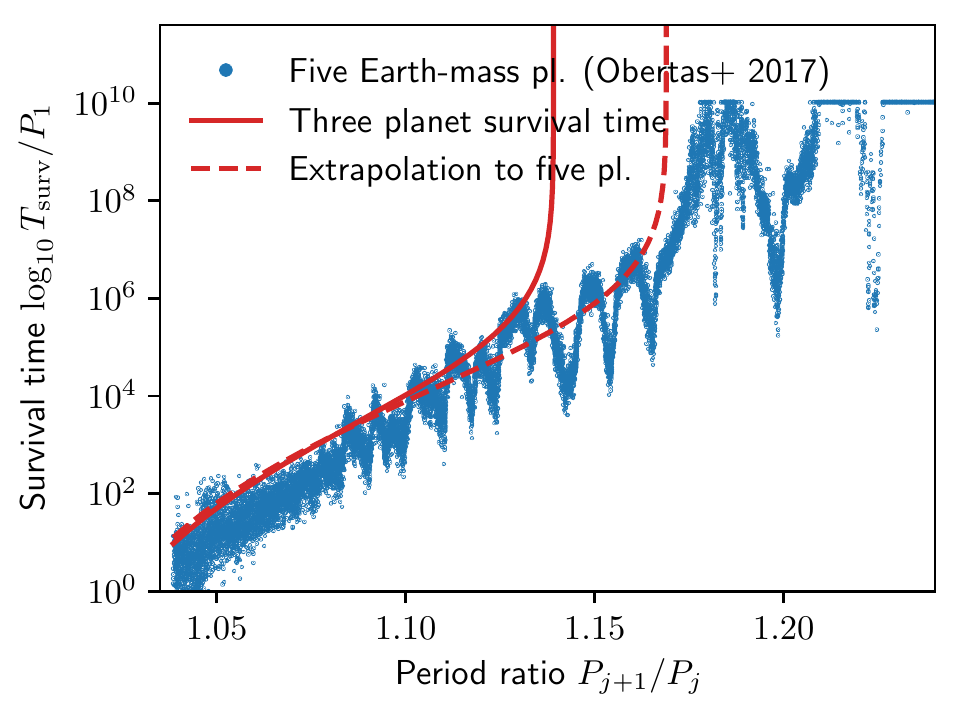}
        \caption{Survival time of five-Earth-mass-planet EMS systems from \citep{Obertas2017} as a function of the period ratio. The red curve corresponds to the survival time of a three-Earth-mass-planet system (eq. \ref{eq:Tsurv_estimate}) and the dashed line is the extrapolation to a five-planet system.\label{fig:obertas}}
\end{figure}

We can compare this expression to previous numerical results.
We choose the simulations from \cite{Obertas2017} as they run systems composed of five Earth-mass planets with high resolution in terms of period ratio.
Moreover, their simulations have been run up to $10^{10}$ orbits.
We plot in Figure \ref{fig:obertas}, the survival time from \cite{Obertas2017} for five-Earth-mass-planet EMS systems as a function of the initial period ratio as well as the three-planet survival time estimate \eqref{eq:Tsurv_estimate} and its extrapolation to a five-planet system.
As expected, the three-planet theoretical survival time slightly overestimates the computed survival time, and more importantly, the MMR overlap criterion fails to account for the continuation of the trend beyond period ratios of 1.14.
On the other hand, the extrapolation to five planets (with $K=2$) goes almost to the region where the survival time starts to increase faster (around 1.165).
We conclude that our approach can successfully account for the difference between three-planet systems and systems of four or more planets.

\subsection{Eccentric planets}

It is tempting to generalise our results to systems with eccentric and inclined planets. 
As noted by \cite{Pu2015}, systems where planets are eccentric have similar survival times to systems with circular orbits if the spacing between the planets is measured by the distance between the apoapsis of the inner planet and the periapsis of the outer planet.
In principle, an overlap criterion could be computed by taking into account the full three-planet MMR network with MMRs of arbitrary order in the same way as was done by \citep{Hadden2018} for the two-planet case.
Another similarity with the two-planet case is the fact that the dynamics of an isolated first-order three-planet MMR are integrable using the same strategy as in the two-planet case (such a result will be presented in a separate article, Petit \emph{in prep.}).

In practice, the structure of the full network is more complicated than the structure of a zeroth-order resonance network and the size of the resonances depends on the individual planet eccentricities.
Moreover, in the case of eccentric orbits, the diffusion is not restricted to a one-dimensional direction.
Indeed one has to take into account the diffusion along the eccentricity degrees of freedom contrarily to the diffusion for circular orbits, because the zeroth-order MMRs conserve the total AMD.
An extension to the eccentric case will be the goal of future works.

\section{Conclusion}

We analyse the mechanism driving the instability time of closely packed planetary systems.
Extending previous work by \cite{Quillen2011}, we use an integrable model to compute the dynamics of three zeroth-order three-planet MMRs for systems on initially circular orbits, with arbitrary planet mass distribution and spacing.
We then compute the region where these resonances overlap (eq. \ref{eq:plsep-ov}), as this is the region where large-scale diffusion can occur \citep{Chirikov1979}.
We find that this region extends past the limit of overlap of two-planet MMR and the spacing scales as $\epsilon^{1/4}$, where $\epsilon$ measures the planet-to-star mass ratio.

Inside the region of overlapping three-planet MMRs, the dynamics are not secular, despite the near-conservation of the AMD, and the period ratios can diffuse up until they reach one of the larger two-planet MMRs, then leading to rapid instability.
We derive an estimated diffusion coefficient by considering only the necessary resonances and as a result, estimate the survival time (eq. \ref{eq:Tsurv_estimate}).
Although in general, Chirikov diffusion leads to survival times that follow power-laws \citep{Quillen2011}, our expression is well approximated by an exponential curve, as is reported in numerical simulations.
Moreover, we predict and observe on numerical simulations a change of behaviour in the region where three-planet MMRs are not overlapping. Beyond the overlap limit, the dynamics cannot be well represented by a relatively uniform diffusion mechanism, and while some other mechanism may destabilise the system, the phenomenon is expected to be much slower.
The stability time therefore depends much more on the initial conditions because other mechanisms such as Arnold diffusion may be necessary to allow the planet to reach the instability.

We compare our results with numerical simulations and find excellent agreement with our analytical estimate.
Moreover, we discuss how apparent discrepancies can be explained.
We also discuss how this result can be extended to systems containing more planets or on eccentric orbits.
Moreover, we show that the classical fit where the instability time is an exponential function of the spacing measured in Hill radius fails to capture the physical mechanism at play.
In particular, we see that for very small bodies, three-body resonances can drive the instabilities over distances that are much larger than single two-planet interactions.
The tools necessary to compute the time estimates and reproduce the figures are made available at \url{https://github.com/acpetit/PlanetSysSurvivalTime}.

In this paper, we focus on systems initially outside of the influence of two-planet MMR.
A similar approach could be applied to the vicinity of a two-planet MMR in order to track the system through the rapid final instability phase.
Such works are necessary to understand the creation of AMD during scattering events, leading to planet collisions and ejections.

\begin{acknowledgements}
	We thank the anonymous reviewer for constructive remarks that improved the manuscript.
	A.P. thanks A. Morbidelli and J. Laskar for useful discussions on the model and M. Pain for providing the references for the random walk exit time.
	We thank A. Obertas and D. Tamayo for allowing us to reproduce their data, and D. Ragozzine, D. Fabrycky and D. T. for their feedback on the preprint.
	This work is by supported by the Royal Physiographic Society of Lund through the Fund of the Walter Gyllenberg Foundation (number 40730).
	M.D. and A.P. are supported by the project grant 2014.0017 ‘IMPACT’ from the Knut and Alice Wallenberg Foundation. A.J. and A.P. are supported by the European Research Council under ERC Consolidator Grant agreement 724687-PLANETESYS, the Swedish Research Council (grant 2018-04867), and the Knut and Alice Wallenberg Foundation (grants 2014.0017 and 2017.0287).
	GP thanks the European Research Council (ERC Starting Grant 757448-PAMDORA) for their financial support.
	This research was made possible by the open-source projects \texttt{rebound} \citep{Rein2012a}, \texttt{Jupyter} \citep{Kluyver2016}, \texttt{iPython} \citep{Perez2007}, \texttt{numpy} \citep{vanderWalt2011}, \texttt{scipy} \citep{Virtanen2020}, \texttt{pandas} \citep{WesMcKinney2010}, and \texttt{matplotlib} \citep{Hunter2007}.
\end{acknowledgements}

\bibliographystyle{aa}
\bibliography{3-planet-instability}

\appendix

\section{Notations summary}
We present in Table \ref{tab:notations} a summary of the notations used in this article.

\begin{table*}
	\centering
	\caption{Summary of the main notations used throughout the article. When possible, we give a short definition and/or refer to the equation where the quantity is defined.\label{tab:notations}}
        \begin{tabular}{c l p{0.5\linewidth} l}
                \hline
                Name & Expression & Description & Reference\vspace{0.1cm}\\
                \hline
                $\vec{r}_j$ & & Heliocentric coordinate position & \cite{Laskar1991} \\
                $\vec{\tilde{r}}_j$ & & Heliocentric coordinate momentum & \cite{Laskar1991} \\
                $\Lambda_j$ & $m_j\sqrt{\mu a_j}$ & Circular angular momentum & \eqref{eq:Delaunay}\\
                $\lambda_j $ & & Mean longitude & \eqref{eq:Delaunay}\\
                $\AMD_j$ & $\Lambda_j\left(1-\sqrt{1-e^2_j}\right)$& Planet $j$  AMD & \eqref{eq:Delaunay}\\
                $\varpi_j$ && Longitude of the periapsis & \eqref{eq:Delaunay}\\
                $x_j$ & $\sqrt{\AMD_j}e^{\i\varpi_j}$ & Complex Poincar\'e coordinate & \eqref{eq:Poincare}\\
                \AMD & $\AMD_1+\AMD_2+\AMD_3$ & Total AMD & \eqref{eq:AM-AMD}\\
                \AM & $\Lambda_1+\Lambda_2+\Lambda_3-\AMD$ & Total angular momentum & \eqref{eq:AM-AMD}\\
                $n_j$ & $\frac{\mu^2m_j^3}{\Lambda_j^3}$ & Mean motion & \eqref{eq:meanmotion}\\
                $\epsilon$ & & Dimensionless parameter related to the planet to star mass ratio &\\
                $\H_0$ & & Keplerian part of the Hamiltonian & \eqref{eq:HKeplerian}\\
                $\epsilon \H_1$ & & Planet interactions Hamiltonian & \eqref{eq:Hpert}\\
                $\epsilon\chi_1$ & & First-order averaging generating Hamiltonian & \eqref{eq:chi}\\
                $\perrat{ij}$ & $P_i/P_j$ & Period ratio for planet $i$ and $j$ & \eqref{eq:perrat}\\
                        $\resloc$ & $\frac{1-\perrat{23}}{\perrat{12}^{-1}-\perrat{23}}$ & Resonance locator & \eqref{eq:resloc}\\
                $\genperrat$ & $\frac{(1-\perrat{12})(1-\perrat{23})}{1-\perrat{12}\perrat{23}}$ & Generalised period ratio separation \\
                \theres & $p \lambda_1 - (p+q)\lambda_2 +q\lambda_3$& Zeroth order three-planet resonant angle & \eqref{eq:transf-angles}\\
                 $\resact$ &$\frac{\Lambda_1}{p}$ & Resonant action & \eqref{eq:transf-actions}\\
                 $\scalefactor$ & $\frac{p+q}{p} \Lambda_1 + \Lambda_2$ & Scaling parameter & \eqref{eq:transf-actions}\\
                \CAM & $\Lambda_1+\Lambda_2+\Lambda_3$ & Circular angular momentum & \eqref{eq:transf-actions}\\
                $\alpha_{ij}$ & $\frac{a_i}{a_j}$ & Semi-major axis ratio & \\
                $ \lapc{s}{l}{\alpha}$ & & Laplace coefficients & \eqref{eq:app-Laplace-coef}\\
                $\epsilon^2 \rescoefpq$ & & Resonant coefficient & \eqref{eq:Rpqexact}, \eqref{eq:Rpqapprox2}\\
                $\plsep_{ij}$ & 1-$\alpha_{ij}$ & Planet orbital spacing & \eqref{eq:plsep-ij}\\
                $\plsep$ & $\frac{\plsep_{12}\plsep_{23}}{\plsep_{12}+\plsep_{23}} $ & Generalised orbital spacing & \eqref{eq:plsep-def}\\        
                $\K_2$ & & Coefficient of the second-order development of the Keplerian part & \eqref{eq:K2}\\
                $\freqpq$ &  $\epsilon\sqrt{\K_2R_{pq}}$ & Small oscillation frequency around the resonance & \eqref{eq:freqpq}\\
                $\numfacfreq$ &  3.47 & Numerical factor appearing in $\freqpq$ & \\
                $\epsilon\Mfac$ & & Relevant mass ratio for the problem & \eqref{eq:Mfac}\\
                $\reseta$ & & Resonance width in period ratio space & \eqref{eq:reswidtheta}\\
                $\density_{p+q}$ & $(p+q)\reseta$ & Density of the subnetwork of resonances with index $p+q$ & \eqref{eq:density_k}\\
                $\denstot$ & & Total density of the zeroth order three-planet MMR network & \eqref{eq:denstot}\\
                $\plsepov$ & & Generalised orbital spacing such that the MMR cover the full space & \eqref{eq:plsep-ov}\\
                $\ovind$ & & Minimum index such that the resonances with lower index locally cover the period ratio space & \eqref{eq:defovind-implicit}\\
                $\xiov$ & $\ovind\plsep$ & & \eqref{eq:xiovexplicit}\\
                $\diffcoef{p+q}$ & $\left(\reseta\right)^2\freqpq/(2\pi)$ & Diffusion coefficient linked to the resonance $p,q$ & \eqref{eq:diffcoef}\\
                \hline
        \end{tabular}
\end{table*}

\section{Laplace coefficients}
\label{app:Laplace coefficients}

The Laplace coefficients appear naturally in the study of planetary systems through the development of the perturbation part in the three-body problem.
The coefficient $\lapc{s}{l}{\alpha}$ corresponds to the $l$-th Fourier coefficient of the function $(1+\alpha^2-2\alpha\cos(\lambda))^{-s}$, \ie
\begin{equation}
\frac{1}{2}\lapc{s}{l}{\alpha} = \frac{1}{2\pi}\int_{-\pi}^\pi \frac{\cos(\lambda)}{\left(1+\alpha^2-2\alpha\cos(\lambda)\right)^s}\d \lambda.\label{eq:app-Laplace-coef}
\end{equation}
There are recurrence relations between them and we refer to \cite{Laskar1995} for a complete description.

One of the challenges of analytical studies of planet dynamics comes from the estimation of the Laplace coefficients.
Indeed, due to the  third Kepler law, $\alpha$ and $l$ are often tied to each other. 
For example, in the study of first-order MMR, it is necessary to compute an approximation for large $l$  of the coefficient $\lapc{s}{l}{\alpha}$ for $\alpha = (1-1/l)^{2/3}$ \citep{Petit2017}.
In other words, the order in which the limits in terms of $\alpha$ and $l$ are taken is relevant.

\cite{Laskar1995} give an alternative expression for the Laplace coefficients, in terms of hypergeometeric functions
\begin{equation}
	\frac{1}{2}\lapc{s}{l}{\alpha} = \frac{\Gamma(s+l)\alpha^l}{\Gamma(s)\Gamma(l+1)} {}_2F_1(s,s+l;l+1;\alpha^2),
	\label{eq:app-Laplace-coef-hypergeom}
\end{equation}
where $\Gamma$ is the Gamma function and ${}_2F_1$ is the Gaussian hypergeometric function.
\cite{Laskar1995} use expression \eqref{eq:app-Laplace-coef-hypergeom} to show that for $\alpha\to 1$, the Laplace coefficients are independent of $l$.
However, we are interested in an estimate where we fix $\alpha$ and make $l$ take larger and larger values.
We cannot therefore use the equivalent they proposed.

\begin{figure}
	\includegraphics[width=\linewidth]{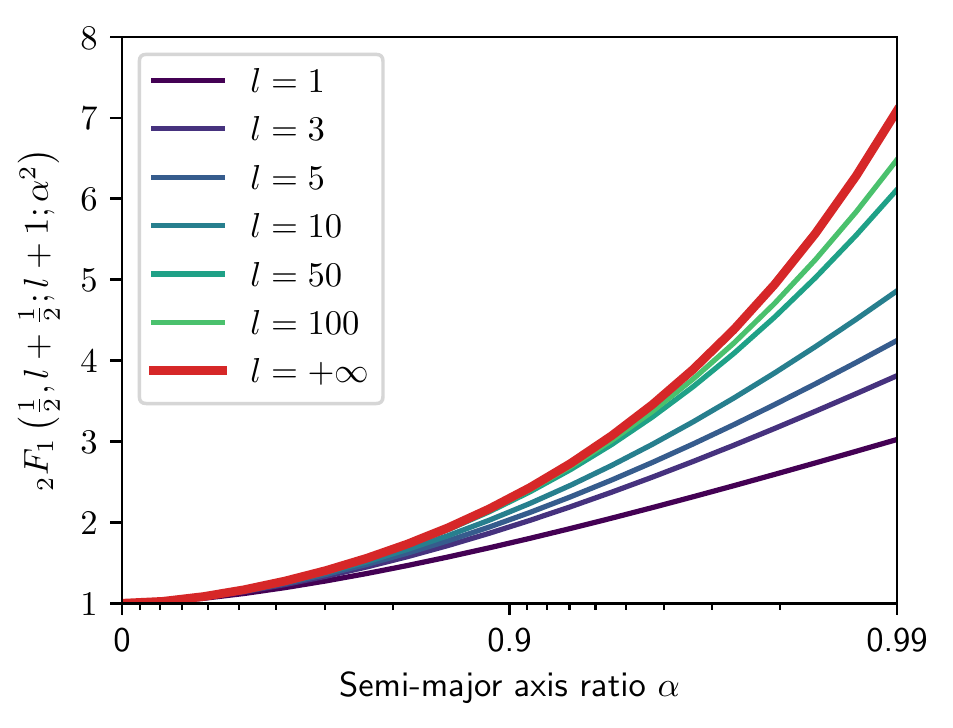}
	\caption{${}_2F_1\left(\frac{1}{2},l+\frac{1}{2};l+1;\alpha^2\right)$ as a function of $\alpha$ for different values of $l$. The case $l=+\infty$ represents the function $f_{1/2}$ given in \eqref{eq:fhalf}\label{fig:hypergeom}}
\end{figure}

In this article, we particularly focus on $\lapc{1/2}{l}{\alpha}$.
For $s=1/2$, Eq. \eqref{eq:app-Laplace-coef-hypergeom} becomes
\begin{equation}
	\frac{1}{2}\lapc{s}{l}{\alpha} = \frac{\alpha^l}{\sqrt{\pi}}\frac{\Gamma\left(l+\frac{1}{2}\right)}{\Gamma(l+1)} {}_2F_1\left(\frac{1}{2},l+\frac{1}{2};l+1;\alpha^2\right).
	\label{eq:app-Laplace-coef-hypergeom1/2}
\end{equation}
In the limit of large $l$, the ratio of $\Gamma$ functions is equivalent to $l^{-1/2}$ (it is worth noting that the estimate is already good for $l=1$).
We can therefore focus on estimating the hypergeometric function.
We find that for $l\to \infty$, ${}_2F_1\left(\frac{1}{2},l+\frac{1}{2};l+1;\alpha^2\right)$ converges to a value depending on $\alpha^2$ that we note $f_{1/2}(\alpha^2)$.
Taking the limit $l$ large into the differential equation verified by the hypergeometric function \citep{Olver2010}\footnote{see \url{https://dlmf.nist.gov/15.10}}, we find that
$f_{1/2}$ is solution of
\begin{equation}
	(1-x)f_{1/2}'(x)-\frac{1}{2}f_{1/2}(x)=0,
\end{equation}
with $f_{1/2}(0)=1$. As a result, we have
\begin{equation}
	f_{1/2}(\alpha^2)=\frac{1}{\sqrt{1-\alpha^2}}.
	\label{eq:fhalf}
\end{equation}
$f_{1/2}(\alpha^2)$ approximates extremely well the hypergeometric function as shown in Figure \ref{fig:hypergeom} where we plot ${}_2F_1\left(\frac{1}{2},l+\frac{1}{2};l+1;\alpha^2\right)$ for different values of $l$ as a function of $\alpha$. We note that $\alpha$ is plotted in logarithmic scale centred on 1 to show where the curve starts to differ.
We observe a fast convergence.

\begin{figure}
	\includegraphics[width=\linewidth]{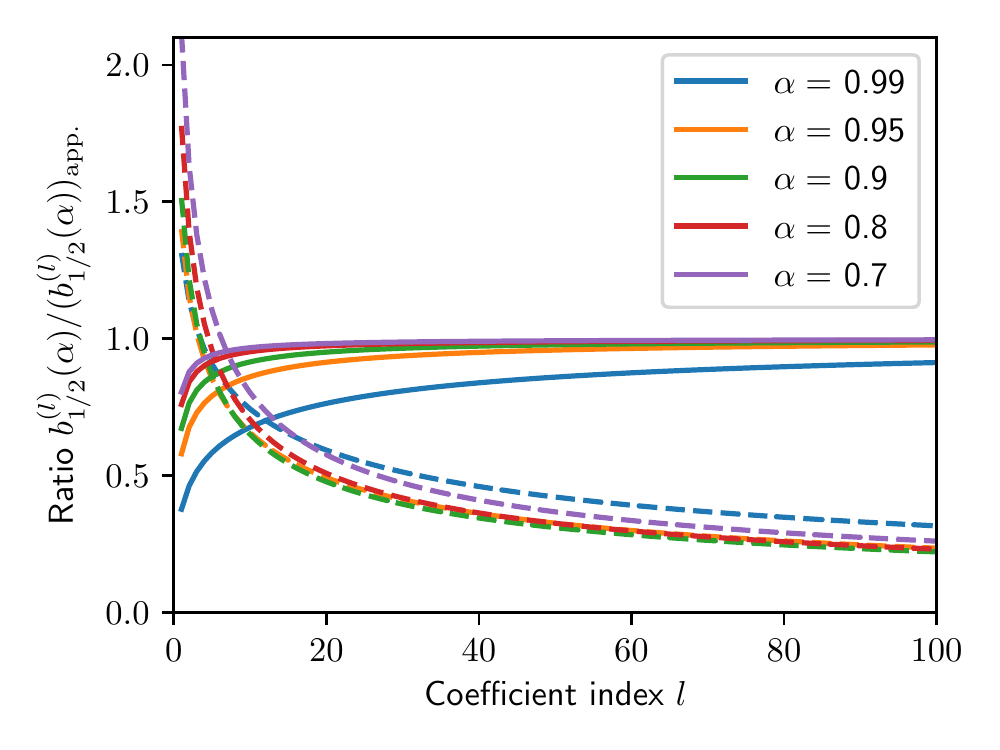}
	\caption{Ratio of the exact Laplace coefficients $\frac{1}{2}\lapc{s}{l}{\alpha}$ with its estimate (\ref{eq:laplaceestimateapp}, full curve) and with Quillen's estimate (dashed curve) as a function of $l$ for different values of $\alpha$.\label{fig-app:laplace}}
\end{figure}

We therefore approximate the Laplace coefficients as
\begin{equation}
	\frac{1}{2}\lapc{s}{l}{\alpha} = \frac{\alpha^l}{\sqrt{\pi l(1-\alpha^2)}}.
	\label{eq:laplaceestimateapp}
\end{equation}
The approximation is very good (within 10\%) for almost all values of $l$.
We plot in figure \ref{fig-app:laplace}, the ratio of the exact Laplace coefficient and its estimate as a function of $l$ for different values of $\alpha$.
In order to compare with \cite{Quillen2011} estimate we plot with dashed line the ratio $\lapc{s}{l}{\alpha}/(|\ln(1-\alpha)|\alpha^l$.
We use $\alpha^l$ instead of $\EXP^{-l(1-\alpha)}$ in Quillen's expression to avoid an unfair comparison since the difference in the exponential would dominate the difference in the prefactors.
It is critical to properly estimate the prefactor.
Indeed, because we integrate over the resonance index in section \ref{sec:network-overlap}, the prefactor contributes significantly to the resonance density, and later to the estimate of the survival time.
As a result, \citep{Quillen2011} estimate fails to fit the Laplace coefficient and as a result, overestimates the resonance width.
In their work, this effect is partially compensated by an underestimation of the Laplace coefficient derivative.

\section{Effective diffusion coefficient estimation}
\label{app:specialfunction}

\subsection{Exact diffusion coefficient}
\label{app:Dawson}

\begin{figure}
\includegraphics[width=\linewidth]{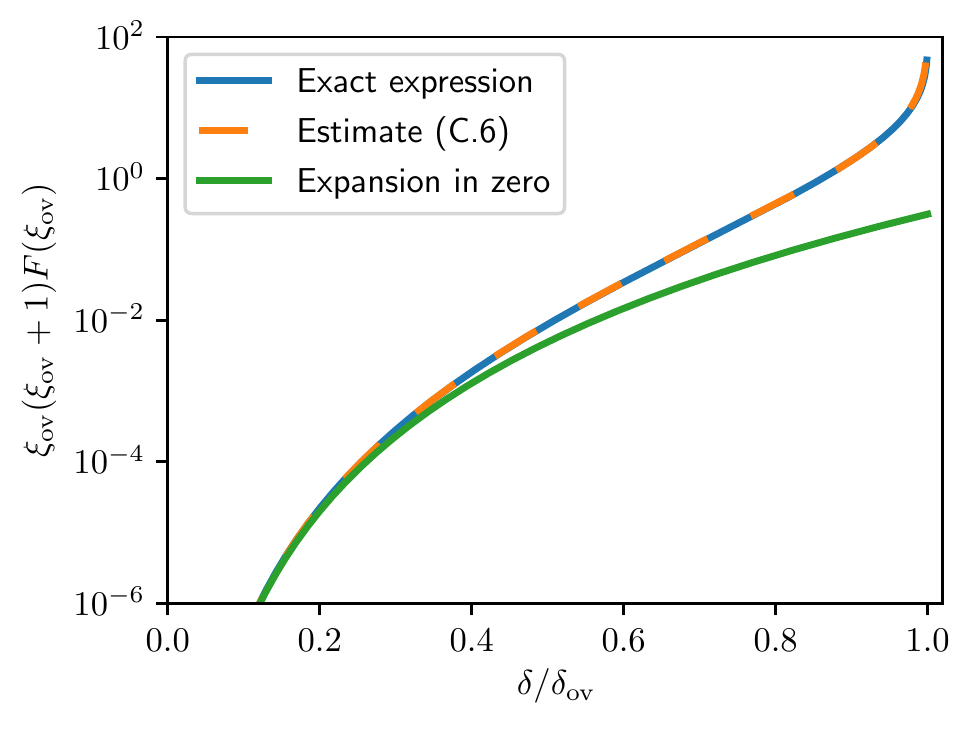}
\caption{$\xiov(\xiov+1)F(\xiov)$ as a function of $\plsep/\plsepov$, its development around zero, and the estimate \eqref{eq:chanceestimate}.\label{fig:Diffcoeffestimate}}
\end{figure}

To compute the effective diffusion coefficient $\diffcoefeff$ (eq. \ref{eq:diffcoef}), one need to solve the integral 
\begin{equation}
	\left(\int_0^{\ovind} \frac{k}{\sqrt{\omega_k}}\d k\right)^{-2} = n_2 \epsilon M \numfacfreq\frac{\sqrt{\resloc(1-\resloc))}}{\plsep} \left(\int_0^{\ovind} \sqrt{k}\EXP^{k\plsep/2}\d k\right)^{-2}\hspace{-0.3cm}.\hspace{-0.3cm}
\end{equation}
The integral can be evaluated in term of the special Dawson function, which gives for $\diffcoefeff$ the expression
\begin{equation}
        \diffcoefeff =\epsilon \Mfac n_2\numfacfreq\sqrt{\resloc(1-\resloc)}\plsep^2\frac{\EXP^{-\xiov}}{\xiov}\left(1-\sqrt{\frac{2}{\xiov}}D_+\left(\sqrt{\frac{\xiov}{2}}\right)\right)^{-2}\hspace{-0.3cm},
	\label{eq:diffcoefex}
\end{equation}
where $\xiov = \ovind\plsep$ is given by Eq. \eqref{eq:xiovexplicit} and $D_+$ is the Dawson function defined as
\begin{equation}
	D_+(x) = \EXP^{-x^2}\int_0^x\EXP^{t^2}dt
.\end{equation}
Using Eq. \eqref{eq:defovind-implicit}, we can replace $\EXP^{-\xiov}$ to obtain
\begin{equation}
        \diffcoefeff =\epsilon \Mfac n_2\numfacfreq\frac{\sqrt{\resloc(1-\resloc)}\plsep^2}{\xiov(\xiov+1)}\left(1-\left(\frac{\plsep}{\plsepov}\right)^4\right)F(\xiov)^{-1},
	\label{eq:diffcoef3}
\end{equation}
with 
\begin{equation}
	F(\xiov) = \left(1-\sqrt{\frac{2}{\xiov}}D_+\left(\sqrt{\frac{\xiov}{2}}\right)\right)^{2}
.\end{equation}
As mentioned in the main text, we find that $\xiov(\xiov+1)F(\xiov)$ is extremely well approximated by 
\begin{equation}
	\xiov(\xiov+1)F(\xiov) \simeq \frac{2\sqrt{2}}{9}\left(\frac{\plsep}{\plsepov}\right)^6 10^{\sqrt{-\ln\left(1-\left(\frac{\plsep}{\plsepov}\right)^4\right)}}.\label{eq:chanceestimate}
\end{equation}
Indeed, the relative difference is below 1\% for $\plsep<0.96$ and within a factor of two overall.
This expression was found by chance after trying to improve an estimation based on a development around zero in terms of $\plsep/\plsepov$.
We plot in figure \ref{fig:Diffcoeffestimate} the exact expression of $\xiov(\xiov+1)F(\xiov)$ as a function of $\plsep/\plsepov$ as well as its estimate. As can be seen, the two curves lie on top of each other.

We also express the numerical factor as a function of $\plsepov$,
\begin{equation}
	\epsilon\Mfac\numfacfreq=\frac{3}{4\sqrt{2}}\frac{\plsepov^4}{(\resloc(1-\resloc))^{3/2}}
.\end{equation}
Combining these terms, we obtain an expression of $\diffcoefeff$ that depends on $\plsep$ and $\plsepov$:
\begin{equation}
	\diffcoefeff \simeq n_2\frac{27}{16}\frac{\plsepov^{10}}{\resloc(1-\resloc)\plsep^4}\left(1-\frac{\plsep^4}{\plsepov^4}\right)10^{-\sqrt{-\ln(1-(\plsep/\plsepov)^4)}}
	\label{eq:diffcoef2}
.\end{equation}

\subsection{Exit time distribution}

We provide here the distribution of the survival time.
As in the main text, the interval where the system can wander has for length $\Delta\resloc = \resloc_{+}-\resloc_{-}$.
The initial position on this interval can be measured by the quantity $u_0= (\resloc_0-\resloc_{-})/\Delta\resloc$ that is between 0 and 1.
The distribution of the log of the survival time $\log_{10}\Tsurv/T_0$ is given by the expression \citep[][eq. 3.0.2]{Borodin2002}
\begin{equation}
	\deriv{P_{\mathrm{surv}}}{\log_{10}\tau} = \sum_{k\in \Z}\frac{(-1)^k(1-u_0+k)}{\sqrt{4\pi \tau}}\exp\left(-\frac{(1-u_0+k)^2}{4\tau}\right),
\end{equation}
where $\tau=\frac{\Tsurv}{T_0}$ and $T_0 = \frac{\Delta\resloc^2}{\diffcoefeff}$.

\end{document}